%% file: sfco.tex
\documentclass[structabstract]{aa} 
\usepackage{graphicx}
\usepackage{txfonts}
\usepackage{natbib}
\usepackage{longtable}
\usepackage{lscape}
\usepackage{multirow}
\usepackage{pdfpages}
\usepackage[colorlinks=true, allcolors=blue]{hyperref}

\bibpunct{(}{)}{;}{a}{}{,}

\newcommand{\kms}{km\,s$^{-1}$}

\newcommand{\degree}{$^{\circ}$}

\begin{document}
\title{Physical and chemical structure of the Serpens filament: fast formation and gravity-driven accretion}

\author{Y. Gong\inst{1}, A. Belloche\inst{1}, F.~J. Du\inst{2}, K.~M. Menten\inst{1}, C. Henkel\inst{1,3}, G.~X. Li\inst{4}, F. Wyrowski\inst{1}, R.~Q. Mao\inst{2}}
\offprints{Y. Gong, \email{ygong@mpifr-bonn.mpg.de, gongyan2444@gmail.com}}

\institute{
Max-Planck-Institut f{\"u}r Radioastronomie, Auf dem H{\"u}gel 69, D-53121 Bonn, Germany
\and 
Purple Mountain Observatory and Key Laboratory of Radio Astronomy, Chinese Academy of Sciences, Nanjing 210034, People's Republic of China
\and
Astronomy Department, Faculty of Science, King Abdulaziz University, P.O. Box 80203, Jeddah 21589, Saudi Arabia
\and
South-Western Institute for Astronomy Research, Yunnan University, Kunming, 650500 Yunnan, People's Republic of China
}

\date{Received date ; accepted date}

\abstract
{The Serpens filament, a prominent elongated structure in a relatively nearby molecular cloud, is believed to be at an early evolutionary stage, so studying its physical and chemical properties can shed light on filament formation and early evolution.}
{The main goal is to address the physical and chemical properties as well as the dynamical state of the Serpens filament at a spatial resolution of $\sim$0.07 pc and a spectral resolution of $\lesssim$0.1~\kms.}
{We performed $^{13}$CO (1--0), C$^{18}$O (1--0), C$^{17}$O (1--0), $^{13}$CO (2--1), C$^{18}$O (2--1), and C$^{17}$O (2--1) imaging observations toward the Serpens filament with the Institut de Radioastronomie Millim{\'e}trique 30-m (IRAM-30 m) and Atacama Pathfinder EXperiment (APEX) telescopes.}
{Widespread narrow $^{13}$CO (2--1) self-absorption is observed in this filament, causing the $^{13}$CO morphology to be different from the filamentary structure traced by C$^{18}$O and C$^{17}$O. Our excitation analysis suggests that the opacities of C$^{18}$O transitions become higher than unity in most regions, and this analysis confirms the presence of widespread CO depletion. Further we show that the local velocity gradients have a tendency to be perpendicular to the filament's long axis in the outskirts and parallel to the large-scale magnetic field direction. The magnitudes of the local velocity gradients decrease toward the filament's crest. The observed velocity structure can be a result of gravity-driven accretion flows. The isochronic evolutionary track of the C$^{18}$O freeze-out process indicates the filament is young with an age of $\lesssim$2 Myr.}
{We propose that the Serpens filament is a newly-formed slightly-supercritical structure which appears to be actively accreting material from its ambient gas.}


\titlerunning{Witnessing filament formation}

\authorrunning{Y. Gong et al.}

\maketitle

\section{Introduction}
Extensive observational work has established that filamentary structures are ubiquitous in the interstellar medium \citep[e.g.,][]{1979ApJS...41...87S,2009ApJ...700.1609M,2010A&A...518L.102A,2010A&A...518L.100M,2010A&A...518L.104M,2013A&A...559A..34L,2016A&A...591A...5L,2020MNRAS.492.5420S,2020A&A...634A.139W}. The properties of these filaments are rather diverse, spanning $\sim$0.1--100 pc in length and $\sim$1--10$^{5}$~$M_{\odot}$ in mass \citep[e.g.,][]{2010A&A...520A..49S,2013A&A...554A..55H,2013A&A...559A..34L,2016A&A...586A..27K,2016A&A...591A...5L,2018A&A...610A..77H,2018A&A...619A.166M,2019A&A...622A..52Z}. A strong correlation between the spatial distribution of dense cores and dense filaments was observed in molecular clouds \citep[e.g.,][]{2015A&A...584A..91K,2016MNRAS.459..342M,2019A&A...631A..72L}. Molecular line observations suggest that material is flowing along filaments to feed nascent stars and clusters as well as dense cores in filament hubs \citep[e.g.,][]{2013ApJ...766..115K,2014A&A...561A..83P,2015ApJ...804...37L,2018ApJ...852...12Y,2018ApJ...855....9L,2018A&A...620A..62G}. In nearby molecular clouds, the filament mass function and filament line mass (mass per unit length) function are consistent with a Salpeter-like mass function, indicating a possible link to the initial mass function \citep{2019A&A...629L...4A}. The core and star formation occurring in filaments may also cause its nearly constant efficiency in dense gas \citep[ $\sim$4.5 $\times 10^{-8}$ yr$^{-1}$,][]{2017A&A...604A..74S}, which in turn regulates the so-called star-formation law in galaxies \citep{2004ApJ...606..271G,2005ApJ...635L.173W}. These results strongly support the idea that filaments play a crucial role in the star-forming processes \citep[e.g.,][]{2010A&A...518L.102A,2014prpl.conf...27A,2015A&A...584A..91K,2017CRGeo.349..187A}. 

Molecular line observations have shown that filaments can consist of bundles of velocity-coherent transonic substructures (``fibers'') with a typical length of $\sim$0.5 pc in both low-mass and high-mass filaments \citep[e.g.,][]{2013A&A...554A..55H,2017A&A...606A.123H,2018A&A...610A..77H,2019A&A...632A..83S,2020ApJ...891...84C}. Such transonic structures can be as long as a few parsecs \citep[e.g.,][]{2016A&A...587A..97H,2018A&A...620A..62G}. Most dense cores (i.e., star-forming seeds) are found in these transonic structures. The formation of transonic structures marks the initial conditions upon which gravitational collapse occurs. Many mechanisms have been proposed to explain filament formation and evolution, which can be divided into three classes of models in which gravity, turbulence, or magnetic fields play a dominant role, respectively \citep[e.g.,][]{2001ApJ...553..227P,2004ApJ...616..288B,2008ApJ...687..354N,2014MNRAS.445.2900S,2014prpl.conf..101L,2014ApJ...791..124G,2018PASJ...70S..53I,2019MNRAS.490.3061V,2019A&A...623A..16S}. 

\begin{figure}[!htbp]
\centering
\includegraphics[width = 0.45 \textwidth]{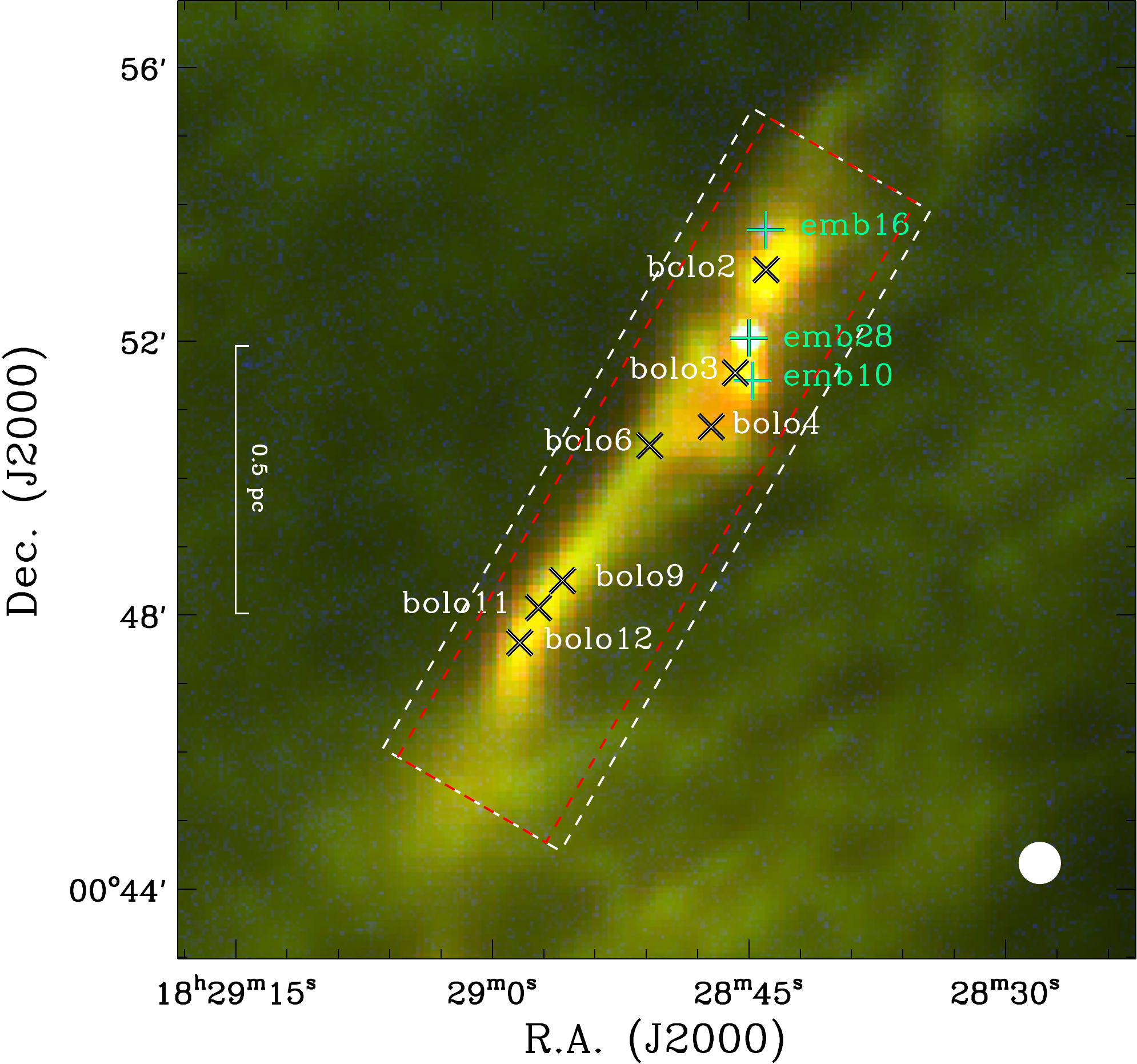}
\caption{{Three-color composite image of the Serpens filament (Herschel 70~$\mu$m: blue, Herschel 250~$\mu$m: green, Herschel 500~$\mu$m: red). The three green pluses indicate the positions of the three embedded YSOs, emb10 (also known as IRAS 18262+0050), emb16, and emb28, \citep{2009ApJ...692..973E}, while the crosses mark the positions of 1.1 mm dust cores \citep{2007ApJ...666..982E}. The beam size (36$\rlap{.}$\arcsec3) for the 500~$\mu$m emission is shown in the lower right corner. The regions mapped with the IRAM-30 m and APEX telescopes are indicated by the red and white dashed boxes, respectively. }\label{Fig:overview}}
\end{figure}

The Serpens filament is a velocity-coherent, transonic filament \citep{2018A&A...620A..62G}. At a distance of $\sim$440 pc \citep{2017ApJ...834..143O,2018ApJ...869L..33O,2019ApJ...879..125Z}, the Serpens filament, which is one of the nearest infrared dark clouds, is a part of 
the Serpens main molecular cloud \citep{2013ApJS..209...39B,2018A&A...620A..62G,2019ApJS..240....9S,2020ApJ...893...91S}. This filament is found to be at the onset of slightly supercritical collapse with a line mass of 36--41~$M_{\odot}$~pc$^{-1}$ and H$_{2}$ column densities of $>$6$\times 10^{21}$~cm$^{-2}$ \citep{2018A&A...620A..62G}. Its relative proximity allows for detailed analyses even with single-dish observations. Figure~\ref{Fig:overview} presents a three-color composite image of the Serpens filament. Three embedded young stellar objects (YSOs) and seven dust cores have been revealed by the 1.1 mm Bolocam continuum and Spitzer legacy surveys \citep{2007ApJ...666..982E,2009ApJ...692..973E}. Follow-up molecular line observations have demonstrated the filament's overall extremely quiescent nature and  also exceptionally slow radial and longitudinal infall motions 
\citep{2018A&A...620A..62G}. This is likely due to a scenario in which most of the filament's material is still at an early evolutionary stage when collapse has just begun. Furthermore, the filament exhibits rather simple geometric and velocity structures, and its southeastern part is truly starless and thus not affected by stellar feedback. Given these characteristics, the Serpens filament lends itself for detailed studies of its 
physical/chemical properties and its underlying dynamical state, which, in turn, should shed light on the general mechanism of filament formation and early evolution. However, the angular resolution of previous molecular line observations ($\sim$1\arcmin, or 0.13 pc) was insufficient to resolve its radial distribution. We thus performed further observations with higher angular resolution ($\sim$30\arcsec, $\sim$0.07 pc), and the outcome is reported in the present paper. 

In Sect. \ref{Sec:obs} we describe our observation and the archival data used in our study. In Sect. \ref{Sec:res} we present the results, which are extensively discussed in Sect. \ref{Sec:dis}. Our findings are summarized in Sect. \ref{Sec:sum}.





\section{Observations and data reduction}\label{Sec:obs}
\subsection{IRAM-30 m and APEX-12 m observations}
We carried out  $^{13}$CO (1--0), C$^{18}$O (1--0), and C$^{17}$O (1--0) observations with the Institut de Radioastronomie Millim{\'e}trique 30-m (IRAM-30 m\footnote{This work is based on observations carried out under project numbers 102-17 and 150-18 with the IRAM 30m telescope. IRAM is supported by INSU/CNRS (France), MPG (Germany) and IGN (Spain).}) telescope during 2018 May 9--11 and 2019 May 16--20. The EMIR dual-sideband and dual-polarization receiver (E090) was used as frontend \citep{2012A&A...538A..89C}, while the autocorrelator VErsatile SPectrometer Assembly (VESPA) spectrometers and Fast Fourier Transform Spectrometers (FFTSs) were simultaneously used as backend. VESPA was employed to achieve high spectral resolution, while the FFTSs were used to cover wider frequency ranges. Each VESPA spectrometer, with a usable bandwidth of $\sim$35.5 MHz, provides 1817 channels, resulting in a channel width of 19.5 kHz, corresponding to 0.052~\kms\,for C$^{18}$O (1--0). Each FFTS provides spectral channels of 50 kHz, translating to a velocity spacing of 0.14~\kms at 112 GHz.

Observations of $^{13}$CO (2--1), C$^{18}$O (2--1), and C$^{17}$O (2--1) were performed with the Atacama Pathfinder EXperiment 12 meter submillimeter telescope (APEX\footnote{This publication is based on data acquired with the Atacama Pathfinder Experiment (APEX). APEX is a collaboration between the Max-Planck-Institut fur Radioastronomie, the European Southern Observatory, and the Onsala Space Observatory.})  \citep{2006A&A...454L..13G} during 2019 April 29--May 1 (project code: M9511A\_103). The PI230 two sideband (2SB) and dual-polarization receiver, built by the Max-Planck-Institute for Radio Astronomy, was used as  frontend, while the facility FFTS was used as backend \citep{2012A&A...542L...3K}. Each of the 4 FFTS modules has a bandwidth of $\sim$4 GHz and provides 65536 channels, which results in a channel width of 61 kHz, corresponding to 0.083~\kms\,for the C$^{18}$O (2--1) line (219.6 GHz). During the IRAM-30 m and APEX observations, we performed on-the-fly (OTF) maps toward the Serpens filament, and the mapping was scanned alternatively along the long and short axes of the filament in order to reduce striping effects in the image restoration. The adopted off position ($\alpha_{\rm J2000}$=18$^{\rm h}$27$^{\rm m}$42$\rlap{.}^{\rm s}$72, $\delta_{\rm J2000}$=00$^{\circ}$44$^{\prime}$54$\rlap{.}^{\prime \prime}$49) was checked to be emission-free in $^{13}$CO (1--0) with a 1$\sigma$ rms noise level of 96 mK at a channel width of 0.14~\kms. 

The standard chopper-wheel method was used to calibrate the antenna temperature \citep{1976ApJS...30..247U}. The calibration was done about every 10 minutes. We converted the antenna temperature, $T_{\rm A}^{*}$, to the main-beam brightness temperature, $T_{\rm mb}$, with the relation $T_{\rm mb}=T_{\rm A}^{*}\eta_{\rm f}/\eta_{\rm mb}$, where $\eta_{\rm f}$ and $\eta_{\rm mb}$ are the forward and main beam efficiency, respectively. These efficiencies are based on the antenna efficiency reports of the IRAM-30 m\footnote{\url{http://www.iram.es/IRAMES/mainWiki/Iram30mEfficiencies}} and APEX \footnote{\url{http://www.apex-telescope.org/telescope/efficiency/}} telescopes. Pointing was checked on nearby continuum sources every hour, and was found to be accurate to within $\sim$3\arcsec. The uncertainties of the absolute flux calibration are smaller than 10\%. A dispersion of $<$5\% between both polarizations is found between the peak intensities of the same molecular line.

Data reduction and analysis were performed with the GILDAS\footnote{\url{http://www.iram.fr/IRAMFR/GILDAS/}} package \citep{2005sf2a.conf..721P}. Convolving with a Gaussian kernel of 1/3 half-power beam width (HPBW), we gridded the raw data into data cubes with regular separations of 5\arcsec\,between two adjacent pixels. A first-order baseline was subtracted from each spectrum. HPBWs after gridding, $\theta_{\rm B}$, typical noise levels, $\sigma$, and channel widths, $\delta \varv$, are given in Table~\ref{Tab:lin}. Velocities are given with respect to the local standard of rest (LSR) throughout this paper.

\subsection{Archival data}\label{Sec:arc}
We make use of the dust temperature and H$_{2}$ column density maps from the Herschel Gould Belt (GB) Survey\footnote{\url{http://www.herschel.fr/cea/gouldbelt/en/index.php}} \citep{2010A&A...518L.102A,2015A&A...584A..91K,2017MmSAI..88..783F}. These maps were derived from spectral energy distribution (SED) fitting to data at four far-infrared bands, that is, 160, 250, 350, and 500~$\mu$m. The data were smoothed to a common angular resolution of 36$\rlap{.}$\arcsec3. Based on the results of \citet{2015A&A...584A..91K}, typical uncertainties of 0.5 K in dust temperature and 10\% in H$_{2}$ column density are assumed in our analysis.

\input{obs}

\section{Results}\label{Sec:res}
\subsection{Overall distribution}\label{sec.abs}
\begin{figure*}[!htbp]
\centering
\includegraphics[width = 0.95 \textwidth]{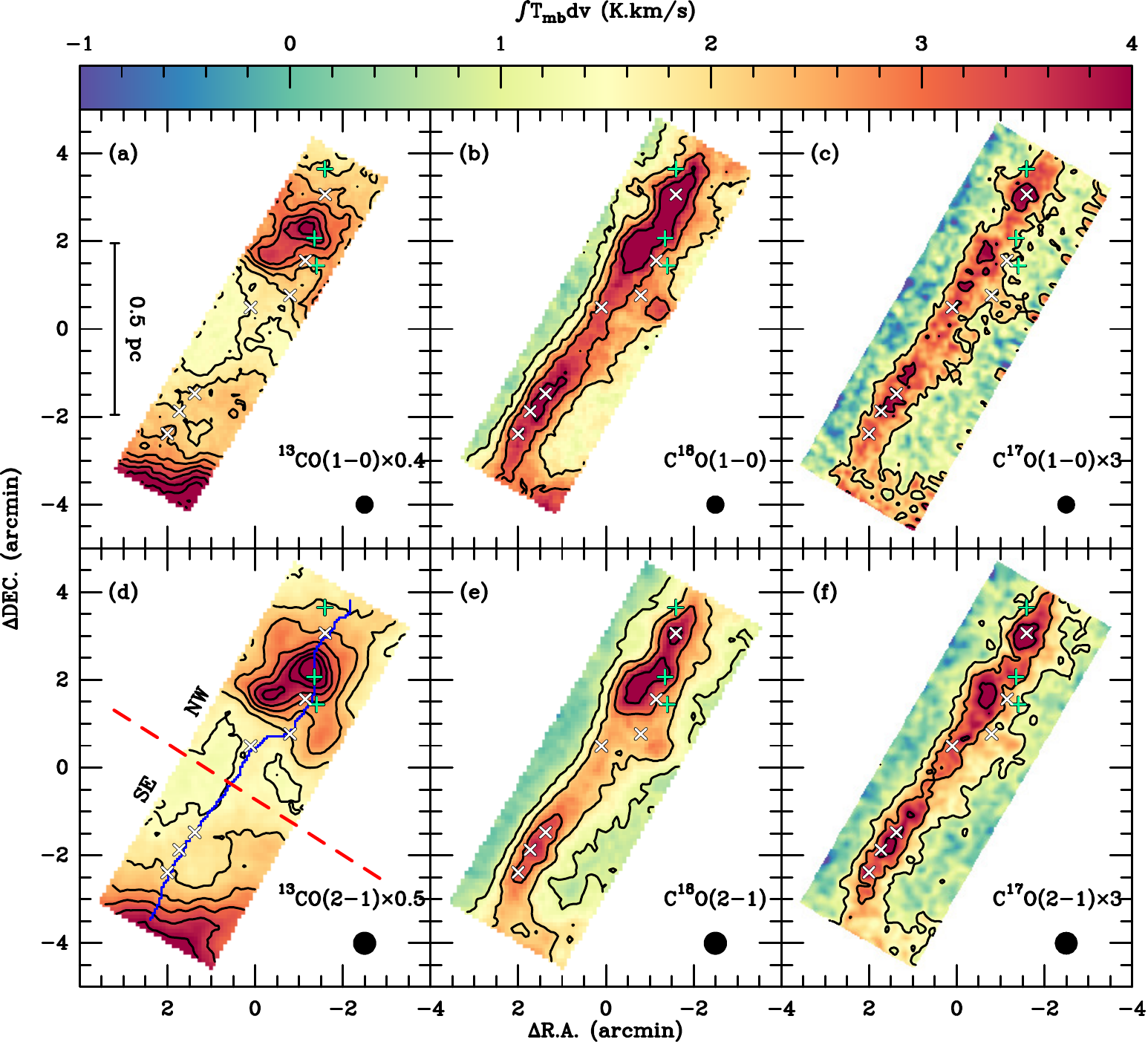}
\caption{{Integrated intensity maps of $^{13}$CO (1--0) (Fig.~\ref{Fig:m0}a), C$^{18}$O (1--0) (Fig.~\ref{Fig:m0}b), C$^{17}$O (1--0) (Fig.~\ref{Fig:m0}c), $^{13}$CO (2--1) (Fig.~\ref{Fig:m0}d), C$^{18}$O (2--1) (Fig.~\ref{Fig:m0}e), and C$^{17}$O (2--1) (Fig.~\ref{Fig:m0}f). Their integrated velocity ranges are [6.5, 10]~\kms\,for $^{13}$CO (1--0), C$^{18}$O (1--0), $^{13}$CO (2--1), and C$^{18}$O (2--1), [7.5, 9]~\kms\,for C$^{17}$O(1--0), and [6, 10]~\kms\,for C$^{17}$O (2--1). The contours start at 0.4~K~\kms with a step of 0.4~K~\kms\,for $^{13}$CO (1--0), C$^{18}$O (1--0), $^{13}$CO (2--1), and C$^{18}$O (2--1). For the two C$^{17}$O maps, the contours start at 3$\sigma$ with a step of 3$\sigma$, where 1$\sigma$=0.08~K~\kms\,for C$^{17}$O (1--0) and 1$\sigma$= 0.06~K~\kms\,for C$^{17}$O (2--1). The color bar represents the C$^{18}$O (1--0) and C$^{18}$O (2--1) integrated intensities in units of K~\kms. The $^{13}$CO (1--0), $^{13}$CO (2--1), C$^{17}$O (1--0), and C$^{17}$O (2--1) maps have been scaled by 0.4, 0.5, 3, and 3 to match the color bar. In Fig.~\ref{Fig:m0}d, the red dashed line is used to divide the Serpens filament into two parts, NW and SE, and the blue line represents the crest of the Serpens filament extracted in the corresponding H$_{2}$ column density map \citep{2018A&A...620A..62G}. In all panels, the (0, 0) offset corresponds to $\alpha_{\rm J2000}$=18$^{\rm h}$28$^{\rm m}$50$\rlap{.}^{\rm s}$4, $\delta_{\rm J2000}$=00$^{\circ}$49$^{\prime}$58$\rlap{.}^{\prime \prime}$72, the three green pluses indicate the positions of the three embedded YSOs, emb10, emb16, and Ser-emb 28 \citep[see][ and Fig.~\ref{Fig:overview}]{2009ApJ...692..973E}, and the white crosses mark the positions of the seven dust cores \citep[see][and Fig.~\ref{Fig:overview}]{2007ApJ...666..982E}.}\label{Fig:m0}}
\end{figure*}

\begin{figure*}[!htbp]
\centering
\includegraphics[width = 0.95 \textwidth]{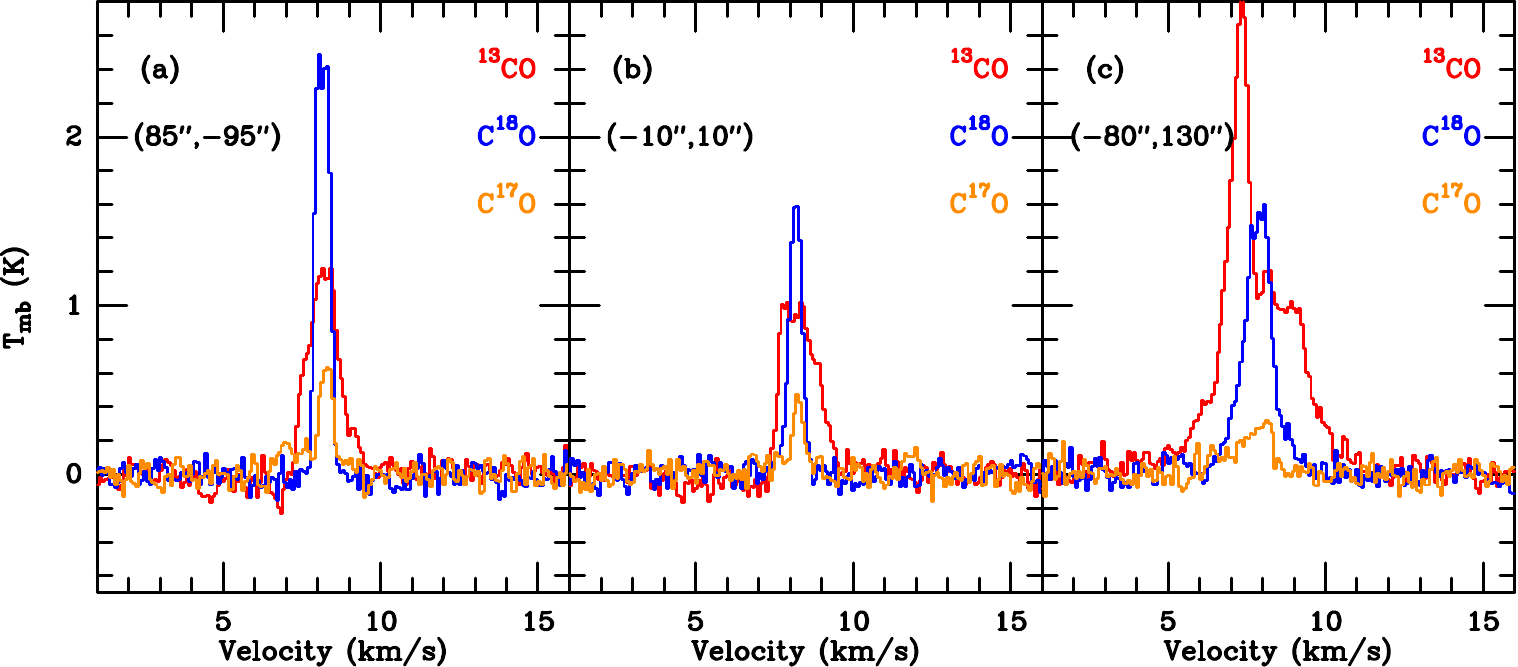}
\caption{{Observed $^{13}$CO (2--1), C$^{18}$O (2--1), and C$^{17}$O (2--1) spectra toward the selected positions that are indicated in the upper left corner. For the origin of the coordinate system, see the caption to Fig.~\ref{Fig:m0}. $^{13}$CO self-absorption is evident in these spectra.}\label{Fig:abs}}
\end{figure*}

Figure~\ref{Fig:m0} presents maps of the integrated intensity of the emission in the $^{13}$CO (1--0), C$^{18}$O (1--0), C$^{17}$O (1--0), $^{13}$CO (2--1), C$^{18}$O (2--1), and C$^{17}$O (2--1) lines resulting from our observations. They provide a factor of two finer linear resolution than our previous C$^{18}$O (1--0) observations \citep{2018A&A...620A..62G}. The red dashed line in Fig.~\ref{Fig:m0}d divides two distinct regions which we call northwest (NW) and southeast (SE) hereafter. The C$^{18}$O and C$^{17}$O line emission shows similar distributions, which trace the filamentary structure well. In contrast, the morphology of the emission of the two $^{13}$CO lines is markedly different. Much of the filamentary structure is not seen in both $^{13}$CO transitions, in line with previous observations \citep{2013ApJS..209...39B}. Also, $^{13}$CO emission is significantly weaker in SE than in NW. Inspecting their corresponding spectra (see Figs.~\ref{Fig:abs}a--\ref{Fig:abs}c and also Appendix~\ref{sec.zoom}), we find that C$^{18}$O (2--1) is generally brighter than $^{13}$CO (2--1) near the systemic velocity of $\sim$8~\kms. This is likely because of significant $^{13}$CO self-absorption, which is reinforced by the presence of the extremely narrow dip in $^{13}$CO spectra at velocities where C$^{18}$O and C$^{17}$O peak. Such narrow $^{13}$CO (2--1) self-absorption features were not reported in previous $^{13}$CO (2--1) observations \citep{2013ApJS..209...39B}, probably due to the coarser spectral resolution of these observations. The widespread $^{13}$CO (2--1) narrow self-absorption suggests high optical depths and lower excitation temperature in the outskirts of the filament or in the ambient gas. 


\subsection{A new molecular outflow driven by Ser-emb 28}\label{sec.outflow}
\begin{figure*}[!htbp]
\centering
\includegraphics[width = 0.9 \textwidth]{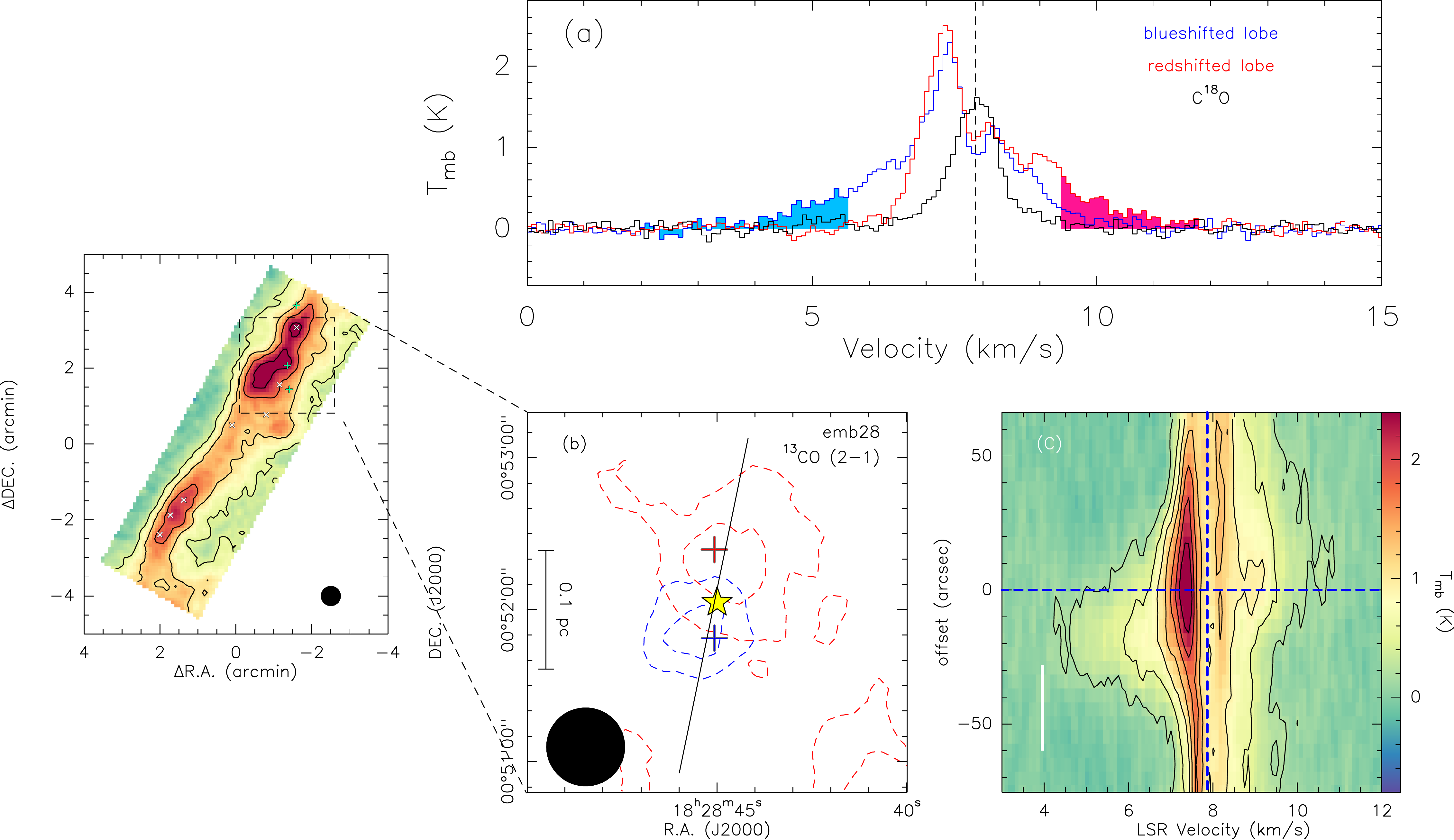}
\caption{{Outflow driven by Ser-emb 28. \textit{Left:} Same as Fig.~\ref{Fig:m0}e. \textit{Right:} (a) Observed $^{13}$CO (2--1) spectra of the two positions indicated by the blue and red pluses in Fig.~\ref{Fig:outflow}b overlaid with the C$^{18}$O spectrum of the position indicated by the yellow star in Fig.~\ref{Fig:outflow}b. (b) $^{13}$CO (2--1) outflow map of Ser-emb 28. The blueshifted emission integrated from 2.0 to 5.5~\kms\,is shown in blue dashed contours which start at 0.20~K~\kms\,(5$\sigma$) and increase by 0.20~K~\kms. The redshifted emission integrated from 9.4 to 12~\kms\,is shown in red dashed contours which start at 0.20~K~\kms\,and increase by 0.20~K~\kms. The yellow star marks the position of Ser-emb 28. The beam size is shown in the lower left corner. (c) Position-velocity diagram of the $^{13}$CO (2--1) emission along the cut indicated by the black line in Fig.~\ref{Fig:outflow}b. The contours start at 0.18~K (3$\sigma$) and increase by 0.36~K. The horizontal dashed line indicates the position of Ser-emb 28, and the vertical dashed line marks its systemic velocity of 7.87~\kms. The  resolution element is shown in the lower left corner.}\label{Fig:outflow}}
\end{figure*}

Based on our APEX $^{13}$CO (2--1) observations which are more sensitive than our $^{13}$CO (1--0) observations, we uncovered a molecular outflow based on its line wing emission (see Fig.~\ref{Fig:outflow}a). Figure~\ref{Fig:outflow}b displays the distribution of the blueshifted and redshifted components which are bluer than 5.5~\kms\,and redder than 9.4~\kms, respectively. Although both blueshifted and redshifted lobes are only marginally resolved, the morphology is suggestive of a bipolar molecular outflow which is nearly oriented in the north-south direction. The outflow direction is at $\lesssim$10\degree\,with respect to the filament. There is an overlap between the blueshifted and redshifted lobes, either caused by our viewing angle or by our limited angular resolution. The YSO Ser-emb 28 is located at this overlap, indicating that the molecular outflow is driven by Ser-emb 28 which was identified as a Class I YSO \citep{2009ApJ...692..973E}. Figure~\ref{Fig:outflow}c presents the position-velocity diagram along the outflow axis. High-velocity wing emission is evident. The brightest blueshifted and redshifted wing emission is located at about 20\arcsec\, (i.e., 8800 au) and 10\arcsec\, (i.e., 4400 au) away from Ser-emb 28, respectively. The blueshifted lobe is found to be stronger than the redshifted lobe. Furthermore, strong $^{13}$CO (2--1) self-absorption is observed around the systemic cloud velocity derived from C$^{18}$O emission. 

The velocity interval of the outflow emission is determined by manually investigating the $^{13}$CO (2--1) channel map (see  Appendix~\ref{append.outflow}).
Assuming that emission in the high velocity wings of the $^{13}$CO (2--1) line is optically thin and is in local thermodynamic equilibrium (LTE) at an excitation temperature of 20 K, we can derive $^{13}$CO column densities in the outflow lobes following \citet{2015PASP..127..266M}. Previous measurements suggest a range of 10--50 K for the excitation temperature in molecular outflows \citep[e.g.,][]{2004A&A...426..503W}. For this range of excitation temperatures, our assumption of a constant 20 K for the excitation temperature would give a $\lesssim$30\% uncertainty in the derived $^{13}$CO  column densities. In order to derive the total molecular mass of the outflow lobes, the $^{13}$CO fractional abundance relative to H$_{2}$ is assumed to be 1.1$\times 10^{-6}$ \citep[e.g.,][]{1992A&A...261..274C,2002A&A...383..892B} and the mean molecular weight per hydrogen molecule is taken to be 2.8 \citep[e.g.,][]{2008A&A...487..993K}. {Arbitrarily}, an inclination angle of 45\degree\,with respect to the line of sight is assumed to correct the outflow parameters. Following previous studies \citep[e.g.,][]{2002A&A...383..892B}, we also determine the momentum, the kinetic energy, the dynamical timescale, the mass entrainment rate, the mechanical force, and the mechanical luminosity. These physical properties are given in Table~\ref{Tab:outflow}. They are comparable to those of outflows from other Class I YSOs \citep[e.g.,][]{2017A&A...600A..99M}. Except for the dynamical timescale, these properties have much lower values than found for outflows in high-mass star-forming regions \citep[e.g.,][]{2018ApJS..235....3Y}. Comparing the outflow's kinetic energy with the turbulent energy of this filament (see Appendix~\ref{append.a}), we find that the kinetic energy contributes $\sim$5\% of the filament's total turbulent energy and $\sim$7\% of NW's turbulent energy. The outflow mechanical luminosity is comparable to the turbulent dissipation rate of this filament (see also Appendix~\ref{append.a}). Hence, the outflow will be able to maintain the observed level of turbulence. We also note that the outflow energy might be dissipated outside the filament, which would reduce the outflow's contribution to the observed turbulence. 

\input{outflow}

\subsection{Decomposition}\label{sec.kine}
\begin{figure*}[!htbp]
\centering
\includegraphics[width = 0.95 \textwidth]{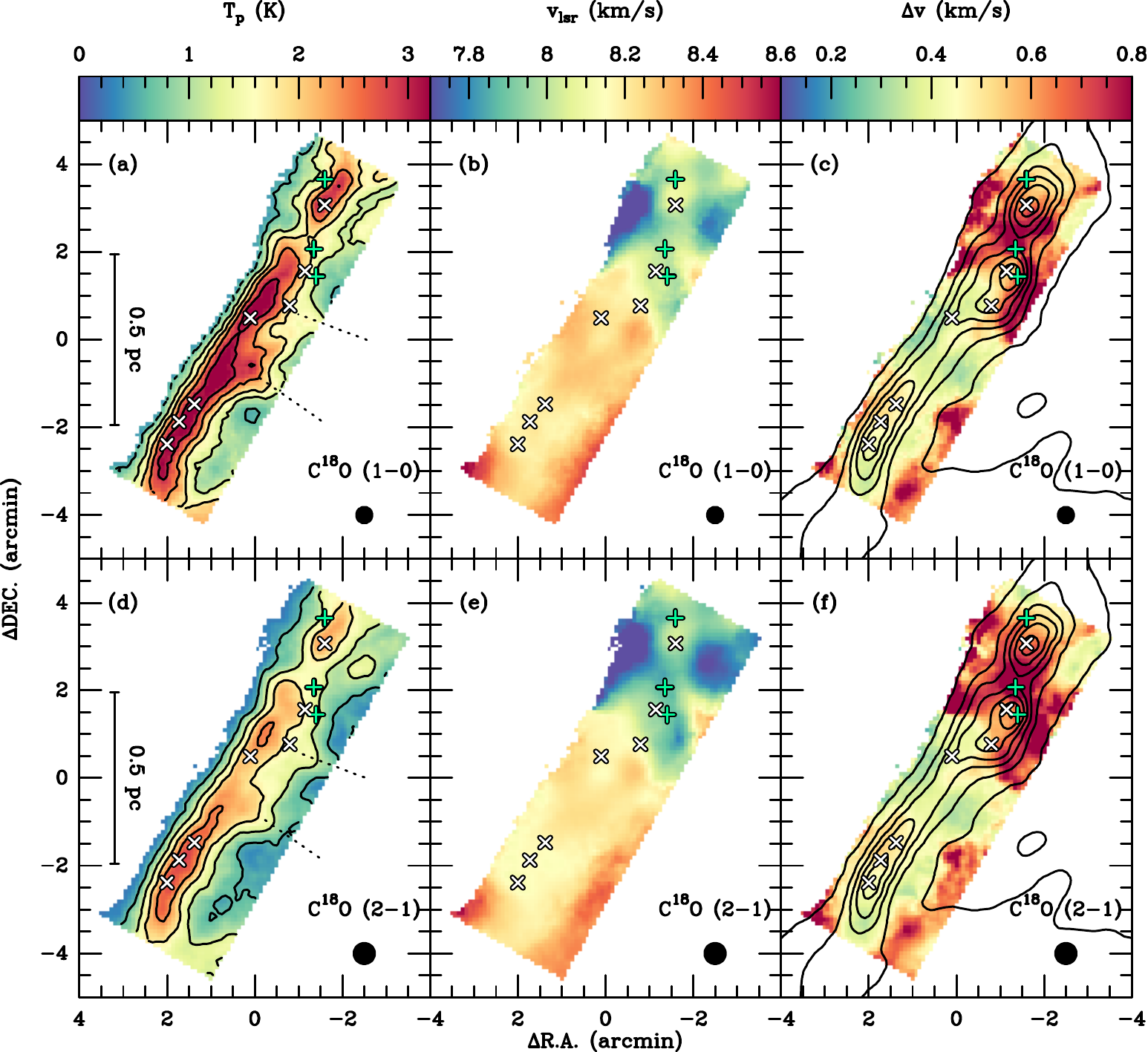}
\caption{{Maps of peak intensities (Fig.~\ref{Fig:kine}a), LSR velocities (Fig.~\ref{Fig:kine}b), and line widths (Fig.~\ref{Fig:kine}c) derived from single-component Gaussian fits to our C$^{18}$O (1--0) data. Figures~\ref{Fig:kine}d--\ref{Fig:kine}f are similar to Figures~\ref{Fig:kine}a--\ref{Fig:kine}c, but for C$^{18}$O (2--1). In Figs.~\ref{Fig:kine}a and \ref{Fig:kine}d, the contours represent the peak intensities which start at 0.6~K and increase by 0.6 K, while in Figs.~\ref{Fig:kine}c and \ref{Fig:kine}f the contours represent the H$_{2}$ column densities that start at 1$\times 10^{21}$~cm$^{-2}$ and increase by 3$\times 10^{21}$~cm$^{-2}$. In Figs.~\ref{Fig:kine}a and \ref{Fig:kine}d, the dotted lines indicate possible substructures. The beam size is shown in the lower right corner of each panel. In all panels, the (0, 0) offset corresponds to $\alpha_{\rm J2000}$=18$^{\rm h}$28$^{\rm m}$50$\rlap{.}^{\rm s}$4, $\delta_{\rm J2000}$=00$^{\circ}$49$^{\prime}$58$\rlap{.}^{\prime \prime}$72, the three green pluses indicate the positions of the three embedded YSOs, emb10, emb16, and Ser-emb 28 \citep{2009ApJ...692..973E}, and the white crosses mark the positions of the seven dust cores \citep{2007ApJ...666..982E}.}\label{Fig:kine}}
\end{figure*}

\begin{figure*}[!htbp]
\centering
\includegraphics[width = 0.95 \textwidth]{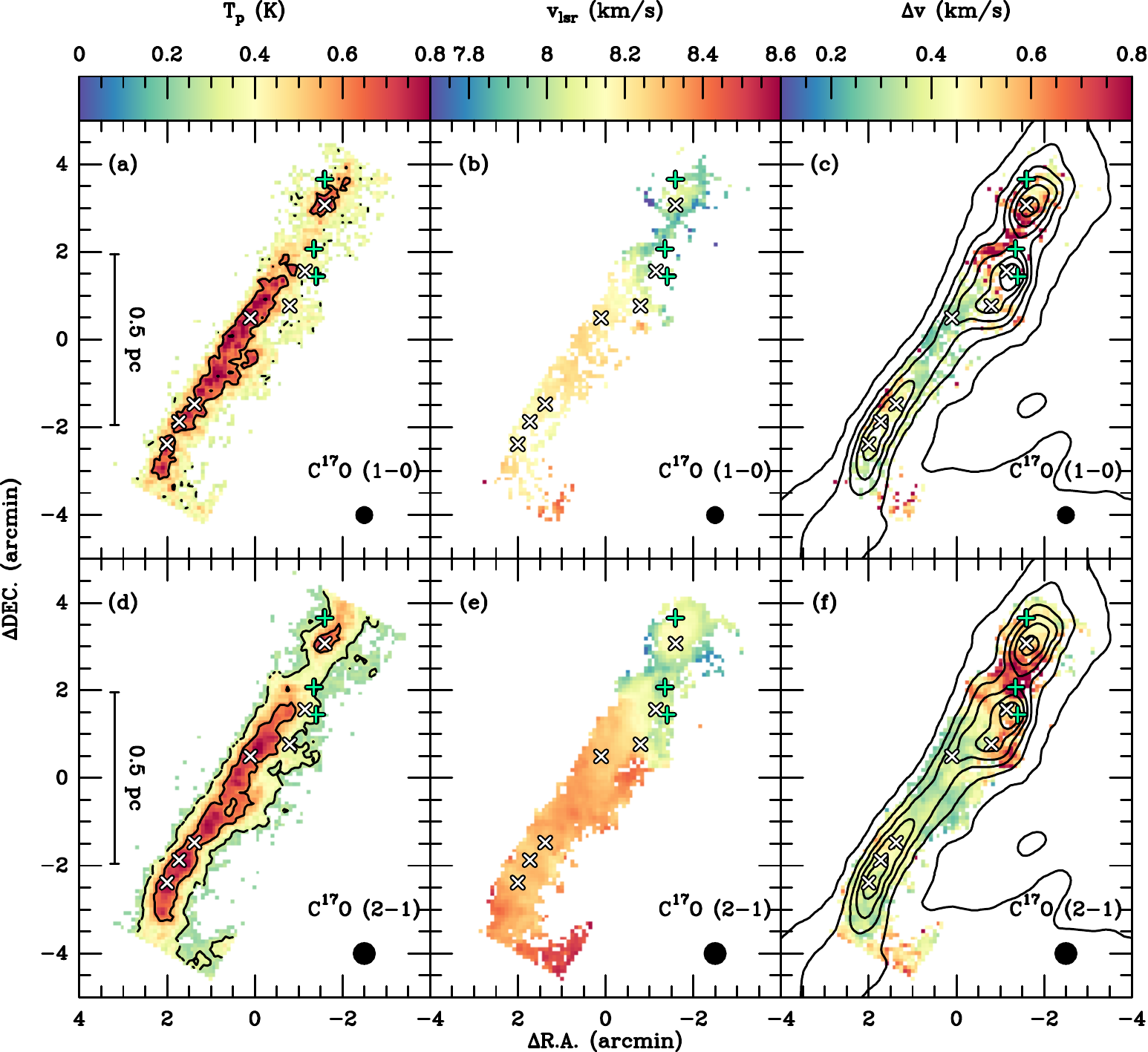}
\caption{{Similar to Fig.~\ref{Fig:kine}, but for C$^{17}$O (1--0) and C$^{17}$O (2--1). In Figs.~\ref{Fig:kine2}a and \ref{Fig:kine2}d, the contours represent the peak intensities which start at 0.3~K and increase by 0.3 K. The pixels with large fitting uncertainties are masked out in Figs.~\ref{Fig:kine2}b--\ref{Fig:kine2}c and Figs.~\ref{Fig:kine2}e--\ref{Fig:kine2}f.}\label{Fig:kine2}}
\end{figure*}

Since $^{13}$CO spectra suffer from self-absorption (see Sect.~\ref{sec.abs}), we only use the C$^{17}$O and C$^{18}$O data to study the kinematics of the Serpens filament. Following previous studies \citep[e.g.,][]{2018A&A...620A..62G}, we apply Gaussian decomposition to the C$^{18}$O (1--0) and C$^{18}$O (2--1) data to obtain their peak intensities, LSR velocities, and line widths for pixels that have at least three adjacent channels with intensities higher than 3$\sigma$. For C$^{17}$O data, which have hyperfine structure (HFS), we apply the HFS fitting routine embedded in GILDAS to decompose them. Due to rather low signal-to-noise ratios, the opacities derived from the HFS fitting have large uncertainties, and are thus not used in the following analysis. The spatial distributions of these observed parameters are displayed in Figs.~\ref{Fig:kine} and \ref{Fig:kine2}.

Figures~\ref{Fig:kine}a and \ref{Fig:kine}d present the peak intensity maps of C$^{18}$O (1--0) and C$^{18}$O (2--1), while Figures~\ref{Fig:kine2}a and \ref{Fig:kine2}d present the peak intensity maps of C$^{17}$O (1--0) and C$^{17}$O (2--1). Similar distributions are evident. The peak intensities are 0.42--3.49~K for C$^{18}$O (1--0),  0.24--2.72~K for C$^{18}$O (2--1),  0.28--0.97~K for C$^{17}$O (1--0), and 0.21--0.87~K for C$^{17}$O (2--1). Assuming an excitation temperature of 7 K (see Sect.~\ref{sec.abund}) and neglecting beam dilution effects, we obtain optical depths of 0.07--2.51 for C$^{18}$O (1--0), 0.09--3.85 for C$^{18}$O (2--1), 0.08--0.30 for C$^{17}$O (1--0), and 0.08--0.38 for C$^{17}$O (2--1). Adopting a lower excitation temperature will result in even higher optical depths. Therefore, opacity effects cannot be neglected in the following analysis. We also find features extending to the west of the filament and directed nearly perpendicular to the filament, indicated by the dotted lines in Figs.~\ref{Fig:kine}a and \ref{Fig:kine}d. These features may be indicative of substructures that provide mass accretion \citep{2013ApJ...766..115K,2013A&A...550A..38P}. However, these features are not evident in the C$^{18}$O integrated intensity and H$_{2}$ column density maps, probably due to their low contrast.  


Figures~\ref{Fig:kine}b, \ref{Fig:kine}e, \ref{Fig:kine2}b, and \ref{Fig:kine2}e show the velocity centroid maps derived from the C$^{18}$O (1--0), C$^{18}$O (2--1), C$^{17}$O (1--0), and C$^{17}$O (2--1) spectra. These plots show global velocity gradients similar to those reported by \citet{2018A&A...620A..62G}. The inner part of SE with $\varv_{\rm lsr}$=8.1--8.3~\kms is blue-shifted relative to its outskirts where values of $\varv_{\rm lsr}$=8.4--8.6~\kms\, are found, whereas NW shows an opposite trend with $\varv_{\rm lsr}$=7.8--8.1~\kms\, in its inner part and $\varv_{\rm lsr}$=7.5--7.9~\kms\, in its outskirts. NW, showing more clumpy structures and more active star formation, is more blueshifted than SE with respect to the ambient gas derived from C$^{18}$O (1--0) at a linear scale of 0.13~pc \citep[$\varv_{\rm LSR}\sim$8.6~\kms,][]{2018A&A...620A..62G}.  


Figures~\ref{Fig:kine}c, \ref{Fig:kine}f, \ref{Fig:kine2}c, and \ref{Fig:kine2}f exhibit line width maps derived from the C$^{18}$O (1--0), C$^{18}$O (2--1), C$^{17}$O (1--0), and C$^{17}$O (2--1) data. Due to their lower signal-to-noise ratios, the fitting results for C$^{17}$O are not as good as those for C$^{18}$O data and for the former isotopologue line width fits with a significance lower than 3$\sigma$ are discarded. We obtain median errors of 0.02~\kms, 0.02~\kms, 0.05~\kms, and 0.04~\kms\,for the fitted line widths of C$^{18}$O (1--0), C$^{18}$O (2--1), C$^{17}$O (1--0), and C$^{17}$O (2--1), respectively. We see that the C$^{18}$O (1--0) and C$^{18}$O (2--1) lines are broader than the C$^{17}$O lines in a significant number of pixels, despite the relatively large uncertainties in the C$^{17}$O fitting results. This difference can be attributed to opacity broadening effects \citep[e.g.,][]{2016A&A...591A.104H}. Corrections for the opacity broadening effect are discussed in Appendix~\ref{append.a}. The opacity broadening corrected velocity dispersion analysis confirms the (tran-)sonic nature of the inner part of the whole filament unveiled by our previous observations \citep{2018A&A...620A..62G}. Such (tran-)sonic nature has already been widely found in both high-mass and low-mass star-forming regions \citep[e.g.,][]{2010ApJ...712L.116P,2016A&A...587A..97H,2017A&A...606A.133S,2020arXiv200313534L}. The opacity-corrected velocity dispersion, $\sigma_{\rm nt,c}$, cannot be derived for low-brightness regions, where the C$^{18}$O opacity is not valid due to low signal-to noise ratios (see Appendix~\ref{append.a}). For consistency, we use non-opacity-corrected velocity dispersions, $\sigma_{\rm nt}$, for discussions unless specified otherwise. Corrections for the low-brightness regions should be small due to their low opacities anyway. Most of the molecular gas has line widths of $<$0.6~\kms\,in the inner region of SE, corresponding to $\sigma_{\rm nt} <$0.25~\kms\,at an assumed kinetic temperature of 7 K. Higher line widths of $>$0.6~\kms\, ($\sigma_{\rm nt}>$0.25~\kms) are observed in NW and the outer region ($N_{\rm H2} \lesssim 10^{22}$~cm$^{-2}$) of SE, suggestive of higher turbulent motions. Generally, NW is more turbulent than SE, probably because NW has more powerful accretion flows. This is supported by larger velocity gradients (see Sect.~\ref{sec.vg}). Furthermore, the higher velocity dispersion around emb28 is also attributed to the outflow feedback because of their spatial coincidence and high turbulent energy contributions (see Sect.~\ref{sec.outflow}).


\subsection{Molecular excitation}\label{sec.abund}
As mentioned above, opacity effects should be taken into account in the analysis. 
According to the radiative transfer equation, the main beam temperature, $T_{\rm mb}$, can be expressed as
\begin{equation}\label{f.rad}
 T_{\rm mb} = \eta (J(T_{\rm ex})-J(T_{\rm bg}))(1-{\rm e}^{-\tau})\;,
\end{equation}
\begin{equation}\label{f.j}
J(T) = \frac{h\nu}{k} \frac{1}{{\rm e}^{\frac{h\nu}{kT}} -1} \;,
\end{equation}
where $\eta$ is the beam dilution factor that is assumed to be unity, $T_{\rm ex}$ is the excitation temperature, $T_{\rm bg}$ is the background temperature that is taken to be 2.73 K, $\tau$ is the optical depth, $h$ is the Planck constant, $k$ is the Boltzmann constant, and $\nu$ is the transition's rest frequency. Because both C$^{17}$O transitions are affected by HFS splitting, we used the $F=$7/2--5/2 component of C$^{17}$O (1--0) and the combined component, Group I (see details in Appendix~\ref{append.b}) of C$^{17}$O (2--1) in Eq.~\ref{f.rad}, and their optical depths are 0.4444 and 0.6933 times the total optical depths of their corresponding total C$^{17}$O transitions. We also note that these components can be overestimated due to overlapping HFS components. This effect can result in up to 28\% and 3\% uncertainties (see details in Appendix~\ref{append.b}). We thus introduce additional 28\%\,and 3\% uncertainties for the peak intensities for the respective component of C$^{17}$O (1--0) and C$^{17}$O (2--1) in the following calculations.

\begin{figure}[!htbp]
\centering
\includegraphics[width = 0.45 \textwidth]{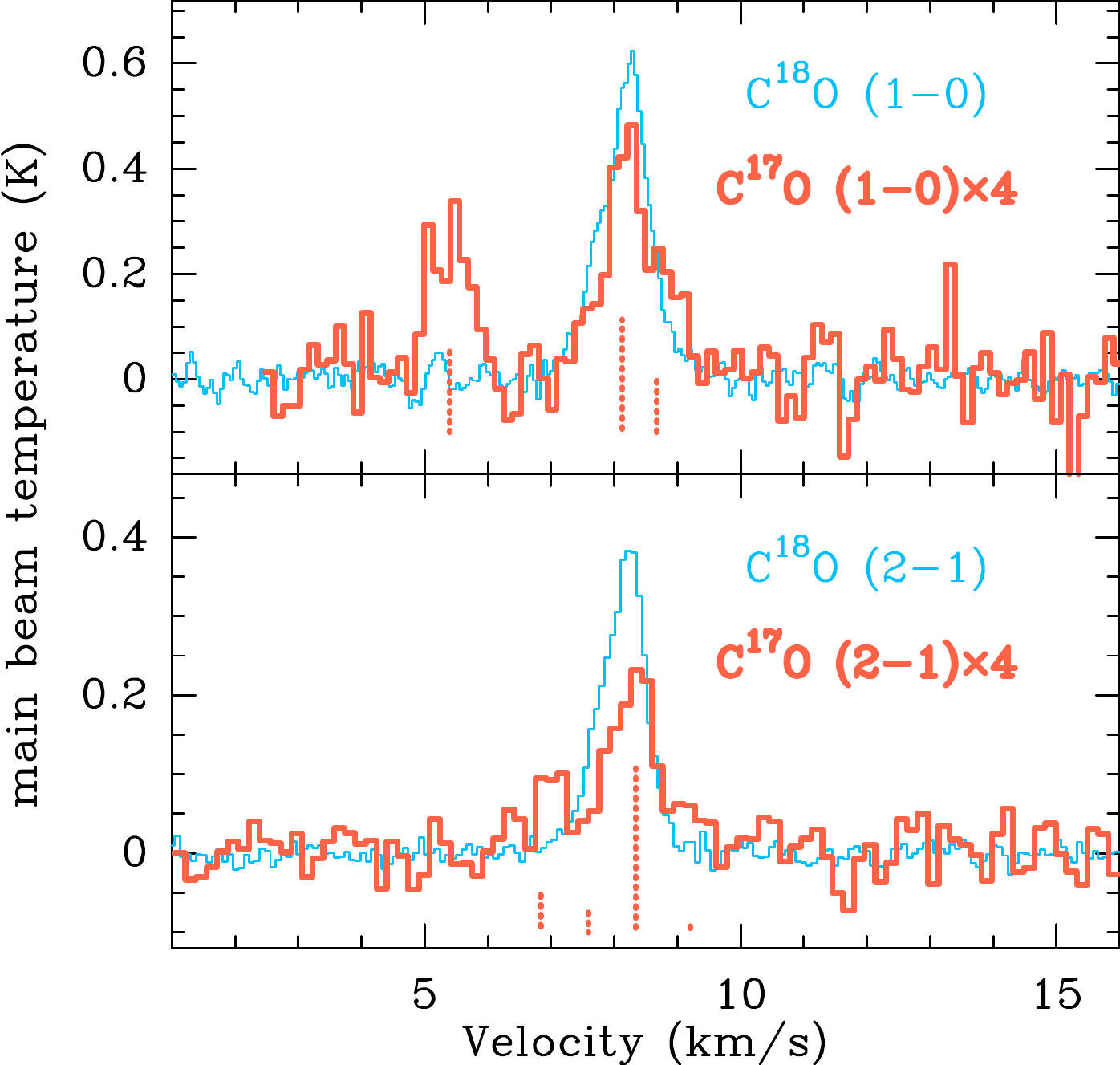}
\caption{C$^{18}$O (1--0), C$^{17}$O (1--0), C$^{18}$O (2--1), and C$^{17}$O (2--1) spectra averaged over the low-brightness-temperature mask (details are discussed in Sect.~\ref{sec.abund}). The two C$^{17}$O spectra have been scaled by a factor of 4. The hyperfine structure components of both C$^{17}$O transitions are indicated by vertical dashed lines. \label{Fig:avsp}}
\end{figure}

In order to constrain molecular excitation, we use the rotational diagram method under the assumption of LTE. By using the optical depth correction factor, $C_{\tau}$=$\frac{\tau}{1-e^{-\tau}}$, the level populations result in the following relation \citep[e.g.,][]{1999ApJ...517..209G}:
\begin{equation}\label{f.rt}
N_{\rm tot} = \frac{3kW}{8\pi^{3}\nu\mu^{2}S} Q(T_{\rm rot}) e^{\frac{E_{\rm u}}{kT_{\rm rot}}} C_{\tau}\;,
\end{equation}
where $\mu$ is the permanent dipole moment of the corresponding molecule, $S$ is the transition's intrinsic strength, $N_{\rm tot}$ is the total molecular column density, $W$ is the integrated intensity of the corresponding transition, $T_{\rm rot}$ is the rotational temperature, $Q$ is the partition function for the corresponding molecule, and $E_{\rm u}$ is the upper level energy of the transition. We took the values of $Q$ and $\mu^{2}S$ from the CDMS \citep{2005JMoSt.742..215M}. We also assume a homogeneous [$^{18}$O/$^{17}$O] isotopic ratio, $r_{\rm O}$, in this filament. This gives us the following relations:
\begin{equation}\label{f.iso}
  r_{\rm O} = N_{18}/N_{17}\;,
\end{equation}
\begin{equation}\label{f.tau}
  r_{\rm O} = \tau_{18, i}/\tau_{17, i}\;,
\end{equation}
where $N_{18}$ and $N_{17}$ are the molecular column densities of C$^{18}$O and C$^{17}$O, and $\tau_{18,i}$ and $\tau_{17,i}$ are the peak optical depths of the corresponding C$^{18}$O and HFS-corrected C$^{17}$O transitions. In Eq.~\ref{f.tau}, we neglect the small difference ($\sim$3\%) between the column density ratios and opacity ratios. In order to make use of Eqs.~\ref{f.iso}--\ref{f.tau}, we first determined the [$^{18}$O/$^{17}$O] isotopic ratio in low brightness regions. For this, we created a mask where the C$^{18}$O (1--0) and C$^{18}$O (2--1) peak intensities are within 0.27--1.0 K and 0.2--0.7 K. This enables the pixels in the mask to have at least 3$\sigma$ while their optical depths are lower than 0.3 at an assumed excitation temperature of 7 K. Therefore, we can assume all C$^{18}$O and C$^{17}$O transitions to be almost optically thin in the mask. In order to improve the signal-to-noise ratios, we averaged the observed spectra in the low-brightness temperature mask and the average spectra are presented in Fig.~\ref{Fig:avsp}. We determined that the integrated intensity ratios $\frac{\int T_{18} {\rm d}\varv}{\int T_{17}{\rm d}\varv}$ are 3.94$\pm$0.23 and 4.41$\pm$0.24 for $J$=1--0 and $J$=2--1, respectively. Using an opacity cut of 0.3 to define the mask means that we may underestimate the true [$^{18}$O/$^{17}$O] isotopic ratio by a factor of 1.15. Using a cut at 0.2 instead would reduce this potential bias. A test shows that the resulting values of the ratios do not change significantly, but the uncertainties increase by a factor of 4. The [$^{18}$O/$^{17}$O] isotopic ratio is thus 3.94--4.41, which is in agreement with previous measurements in other nearby molecular clouds \citep{2005A&A...430..549W,2007A&A...465..887Z,2008A&A...487..237W,2014A&A...570A..65G,2020arXiv200700361Z}. We therefore use a constant $r_{\rm O}$ value of 4.1 in Eqs.~\ref{f.iso}--\ref{f.tau}.


\begin{figure*}[!htbp]
\centering
\includegraphics[width = 0.8 \textwidth]{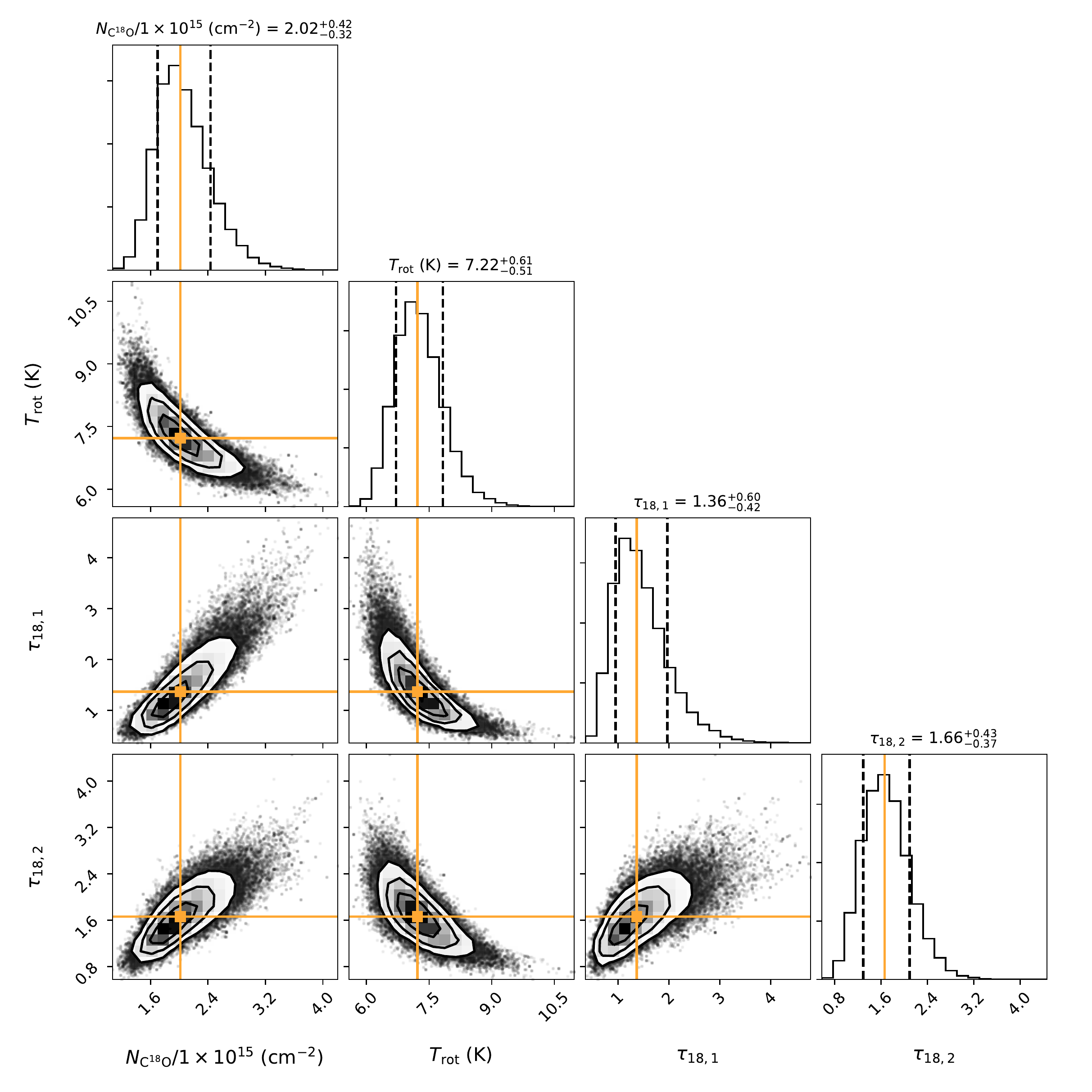}
\caption{{Posterior probability distributions of C$^{18}$O column density, $N_{\rm C^{18}O}$, rotational temperature, $T_{\rm rot}$, the C$^{18}$O (1--0) opacity, $\tau_{18,1}$, and the C$^{18}$O (2--1) opacity, $\tau_{18,2}$, toward the dense core, Bolo12, with the maximum posterior possibility point in the parameter space shown in orange lines and points. Contours denote the 0.5$\sigma$, 1.0$\sigma$, 1.5$\sigma$, and 2.0$\sigma$ confidence intervals.}\label{Fig:ltemcmc}}
\end{figure*}

 In order to solve Eqs.~(\ref{f.rad})--(\ref{f.tau}) for both molecules simultaneously, we assume that all C$^{18}$O and C$^{17}$O transitions have the same excitation (or rotational) temperature. As already mentioned above, we fixed the $r_{\rm O}$ value to be 4.1. For the purpose of improving the signal-to-noise ratios and the following comparison with the dust temperature and dust-based H$_{2}$ column density maps, we first convolved C$^{18}$O and C$^{17}$O peak and integrated intensity maps to the angular resolution of the Herschel maps (i.e., 36$\rlap{.}$\arcsec3). For pixels with signal-to-noise ratios higher than 3, we solved Eqs.~(\ref{f.rad})--(\ref{f.tau}) with the emcee\footnote{\url{https://emcee.readthedocs.io/en/stable/}} code \citep{2013PASP..125..306F} to perform the Monte Carlo Markov chain (MCMC) calculations with the affine-invariant ensemble sampler \citep{2010CAMCS...5...65G} in order to obtain the physical parameters and their uncertainties simultaneously. We assumed uniform priors for $N_{\rm C^{18}O}$, $T_{\rm rot}$, $\tau_{18,1}$, and $\tau_{18,2}$. The posterior distribution of these parameters is given by the product of the prior distribution function and the likelihood function. The likelihood function is assumed to be $\propto {\rm e^{-\chi^{2}/2}}$ with
\begin{equation}\label{f.chi}
  \chi^{2}=\Sigma_{i} (I_{\rm obs, i}-I_{\rm mod, i})^{2}/\sigma_{\rm obs, i}^{2}\;,
\end{equation}
where $I_{\rm obs,i}$ and $I_{\rm mod,i}$ are observed and modeled parameters including peak and integrated intensity, and $\sigma_{\rm obs,i}$ is the standard deviation of $I_{\rm obs,i}$. The MCMC simulations were run with 20 walkers and for 4500 steps after the burn-in period. The 1$\sigma$ uncertainties are given by the 16th and 84th percentiles of the posterior distribution. An example of the fitting result is shown in Fig.~\ref{Fig:ltemcmc}.

\begin{figure}[!htbp]
\centering
\includegraphics[width = 0.45 \textwidth]{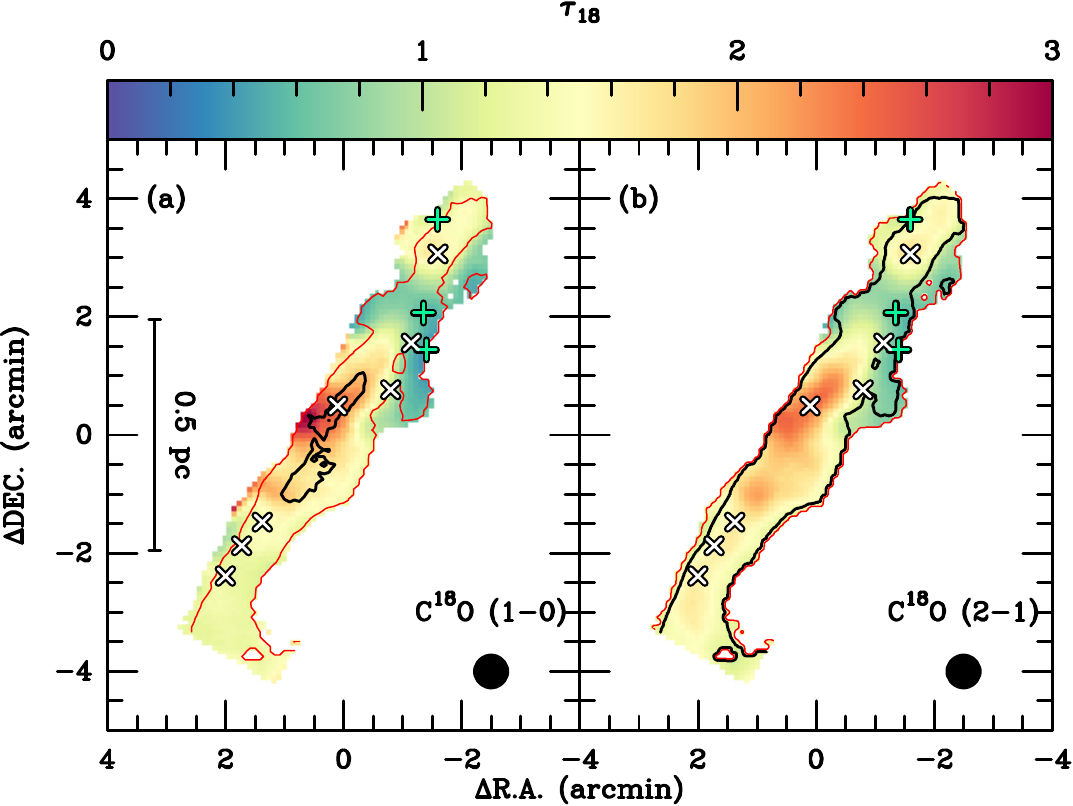}
\caption{{Optical depth maps of C$^{18}$O (1--0) (Fig.~\ref{Fig:tau}a) and C$^{18}$O (2--1) (Fig.~\ref{Fig:tau}b) overlaid with contours of signal-to-noise ratio ($\tau/\sigma_{\tau}$) of 2 and 3 indicated by the red and black lines, respectively. The beam size is shown in the lower right corner of each panel. In all panels, the (0, 0) offset corresponds to $\alpha_{\rm J2000}$=18$^{\rm h}$28$^{\rm m}$50$\rlap{.}^{\rm s}$4, $\delta_{\rm J2000}$=00$^{\circ}$49$^{\prime}$58$\rlap{.}^{\prime \prime}$72, the three green pluses show the positions of the three embedded YSOs, emb10, emb16, and Ser-emb 28 \citep{2009ApJ...692..973E}, and the white crosses mark the positions of the seven dust cores \citep{2007ApJ...666..982E}.}\label{Fig:tau}}
\end{figure}

\begin{figure}[!htbp]
\centering
\includegraphics[width = 0.45 \textwidth]{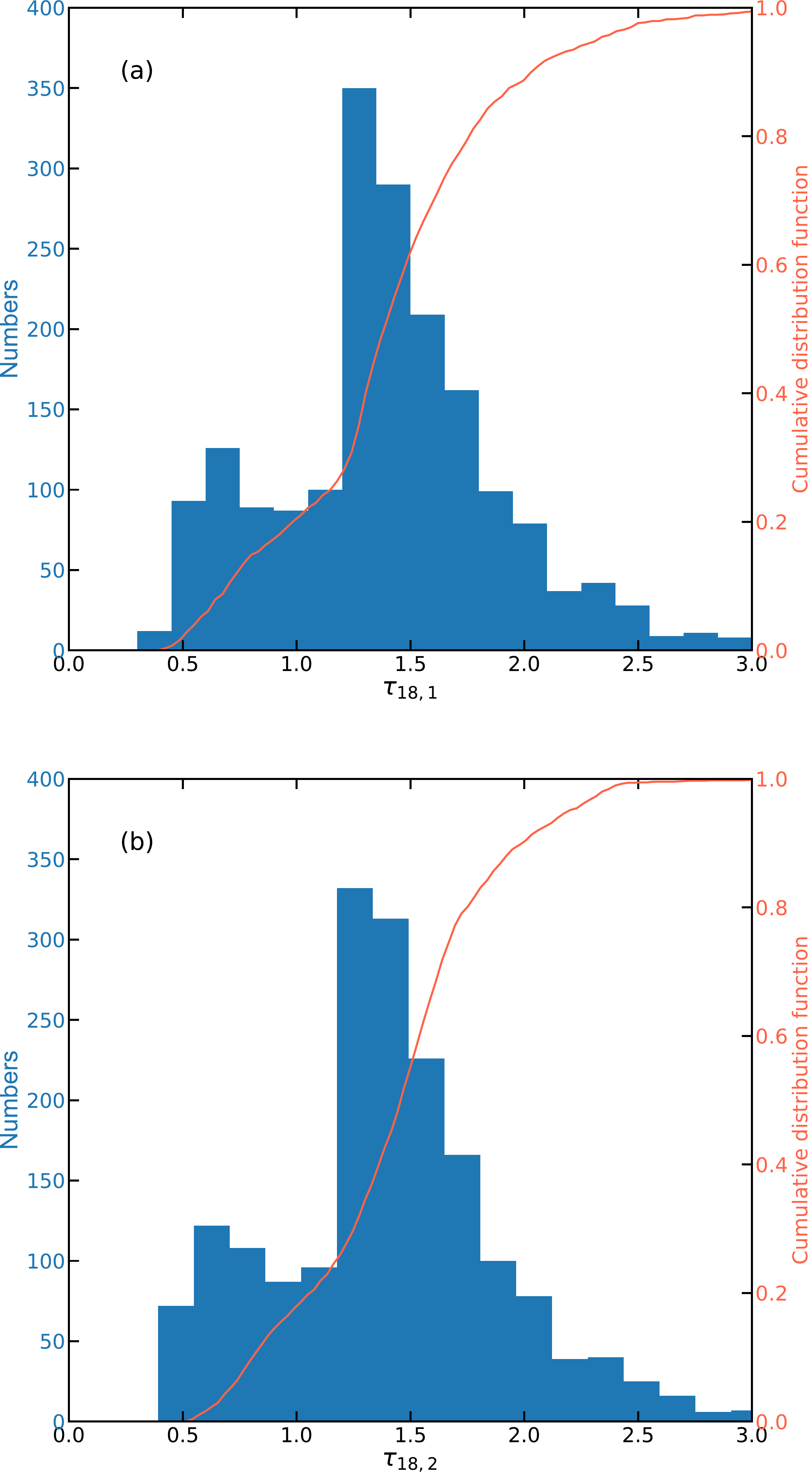}
\caption{{Histogram and cumulative distribution of the C$^{18}$O (1--0) (Fig.~\ref{Fig:histtau}a) and C$^{18}$O (2--1) (Fig.~\ref{Fig:histtau}b) opacities.}\label{Fig:histtau}}
\end{figure}

The optical depth maps of C$^{18}$O (1--0) and C$^{18}$O (2--1) are shown in Fig.~\ref{Fig:tau}. The C$^{18}$O (1--0) opacities lie in the range of 0.3--2.9, while the C$^{18}$O (2--1) opacities are 0.4--2.5. It is surprising that we found more than 80\% of pixels with the C$^{18}$O (1--0) and C$^{18}$O (2--1) optical depths higher than unity (see Fig.~\ref{Fig:histtau}), because these C$^{18}$O lines are usually assumed to be optically thin in the literature \citep[e.g.,][]{2009ApJ...705L..95I,2016A&A...588A.104G,2018A&A...620A..62G}. The highest opacities are found around Bolo6. Such high C$^{18}$O opacities of $>$1 were also reported in other infrared dark clouds \citep[e.g.,][]{2019MNRAS.490.4489S}. 

\begin{figure*}[!htbp]
\centering
\includegraphics[width = 0.95 \textwidth]{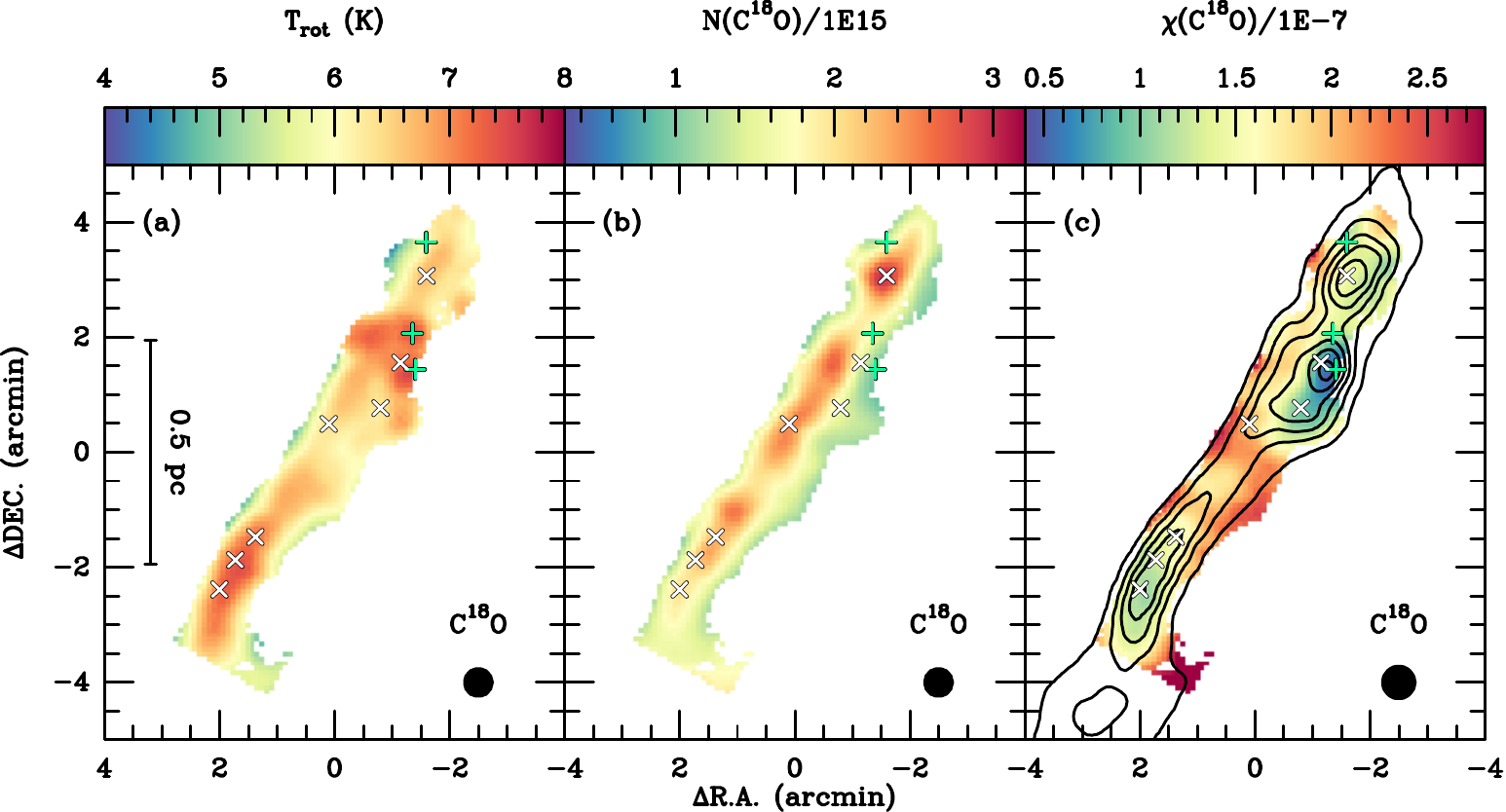}
\caption{{(a) C$^{18}$O rotational temperature map. (b) C$^{18}$O column density map. (c) C$^{18}$O fractional abundance map overlaid with H$_{2}$ column density contours. The contours start at 6$\times 10^{21}$ cm$^{-2}$, and increase by 3$\times 10^{21}$ cm$^{-2}$. The beam size is shown in the lower right corner of each panel. In all panels, the (0, 0) offset corresponds to $\alpha_{\rm J2000}$=18$^{\rm h}$28$^{\rm m}$50$\rlap{.}^{\rm s}$4, $\delta_{\rm J2000}$=00$^{\circ}$49$^{\prime}$58$\rlap{.}^{\prime \prime}$72, the three green pluses indicate the positions of the three embedded YSOs, and the white crosses mark the positions of the seven dust cores \citep{2007ApJ...666..982E}.}\label{Fig:abund}}
\end{figure*}

\begin{figure*}[!htbp]
\centering
\includegraphics[width = 0.95 \textwidth]{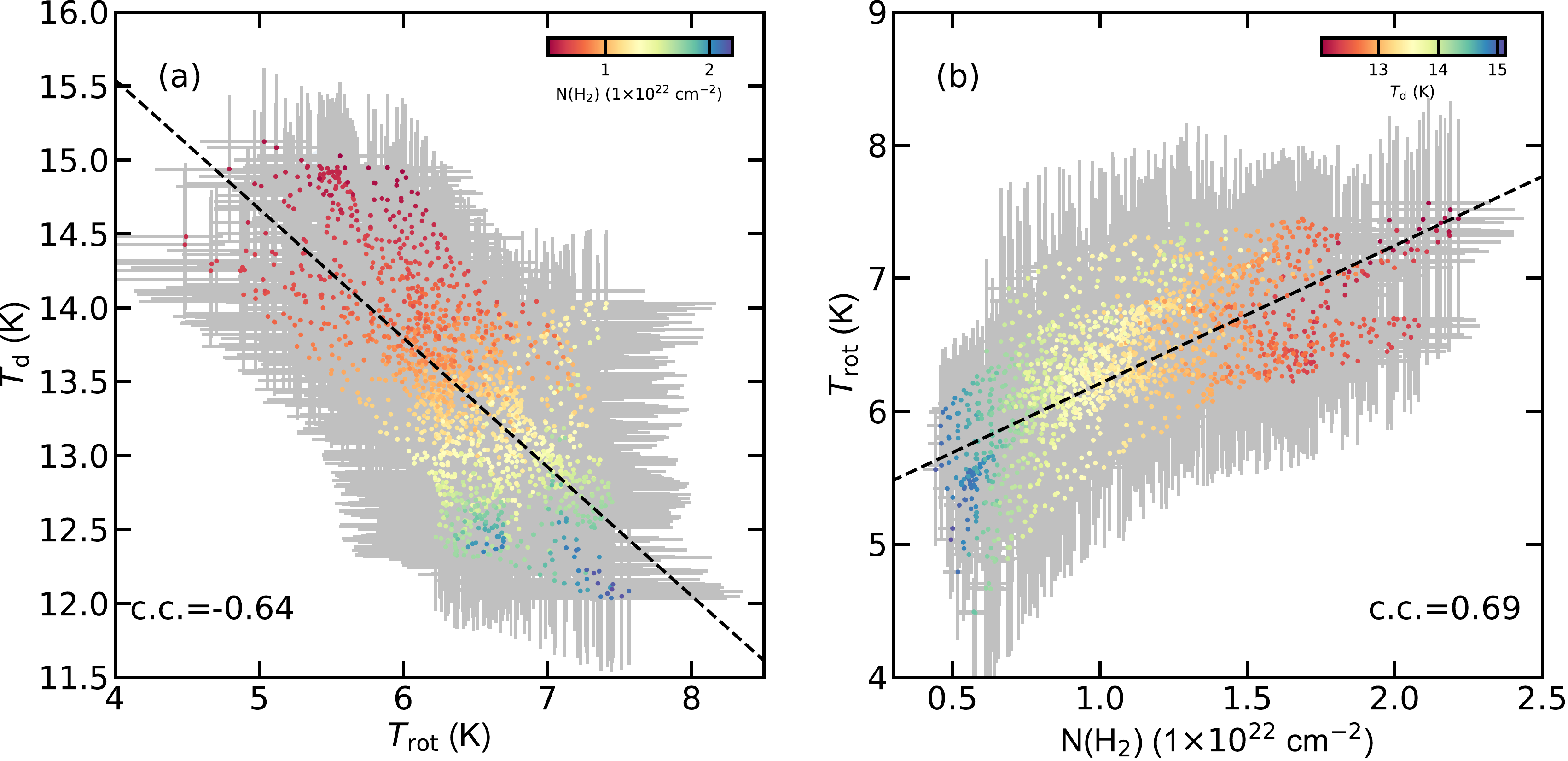}
\caption{{ (a) Dust temperature as a function of the C$^{18}$O rotational temperature. The gray error bars represent the measured uncertainties in the measured dust temperature and rotational temperature. The markers' colors correspond to their H$_{2}$ column densities indicated by the color bar. The Pearson correlation coefficient is shown in the lower left. (b) C$^{18}$O rotational temperature as a function of the H$_{2}$ column density. The gray error bars represent the measured uncertainties in the measured rotational temperature and H$_{2}$ column density. The markers' colors correspond to their dust temperatures indicated by the color bar. The Pearson correlation coefficient is shown in the lower right. In both panels, the black dashed line represents the linear fit to the observed trend.}\label{Fig:comp_t}}
\end{figure*}

Figure~\ref{Fig:abund}a shows the distribution of the rotational temperature that spans 4.6--7.6 K with a median of 6.2~K. All rotational temperatures are lower than 8~K. A pixel-by-pixel comparison between the dust temperature and the rotational temperature is shown in Fig.~\ref{Fig:comp_t}a. The dust temperature is generally higher than the rotational temperature, and decreases with increasing rotational temperatures. A linear fit results in the empirical relation, $T_{\rm d}\;(K) = (-0.8\pm 0.1)T_{\rm rot}\;(K)+(18.5\pm 0.2)$. Theoretical studies suggest that the gas temperature is comparable to or slightly higher than the dust temperature at an H$_{2}$ density of $>10^{4.5}$~cm$^{-3}$ due to the gas-dust coupling, while the gas temperature becomes much higher than the dust temperature at lower H$_{2}$ densities due to the relatively inefficient gas cooling \citep{2001ApJ...557..736G}. The observed empirical relation appears to be inconsistent with the theoretical prediction. The reasons can be: (1) the background and foreground contributions to the dust emission can lead to a higher dust temperature than gas temperature from the SED fitting approach. This is supported by previous radiative transfer simulations \citep[e.g.,][]{2012A&A...547A..11N}; (2) C$^{18}$O and C$^{17}$O transitions become sub-thermally excited in low-density regions ($n_{\rm H2} \lesssim 10^{3}$ cm$^{-3}$), leading to the result that the rotational temperature is lower than the corresponding kinetic temperature. The latter effect becomes more prominent in the lower-density regime, leading to the observed anticorrelation in Fig.~\ref{Fig:comp_t}a. This is also supported by the result that rotational temperatures correlate with H$_{2}$ column densities and high rotational temperatures are found to be associated with high H$_{2}$ column densities (see Fig.~\ref{Fig:comp_t}b). We performed a linear fit to Fig.~\ref{Fig:comp_t}b, and obtain an empirical relation of $T_{\rm rot}\;({\rm K})=(1.04\pm 0.03) \frac{N({\rm H}_{2})}{1\times 10^{22} {\rm cm^{-2}}}+(5.17\pm 0.03)$. The fact that the rotational temperature is lower than the dust temperature in high H$_{2}$ column density regions can also be partially caused by CO depletion (see discussions below). Because the densest regions with higher rotational temperatures cannot be probed by CO due to its depletion, the measured rotational temperatures come from less dense regions that have lower rotational temperatures.


Figure~\ref{Fig:abund}b presents the distribution of C$^{18}$O column densities\footnote{Since the [$^{18}$O/$^{17}$O] isotopic ratio is fixed at 4.1, the distribution of C$^{17}$O column densities is the same as that of C$^{18}$O but with lower values.}. C$^{18}$O column densities range from 9.1$\times 10^{14}$ cm$^{-2}$ to 3.0$\times 10^{15}$ cm$^{-2}$ with a median of 1.7$\times 10^{15}$~cm$^{-2}$, which almost doubles the values reported in \citet{2018A&A...620A..62G}. This is because \citet{2018A&A...620A..62G} did not take C$^{18}$O opacity effects into account and assumed a slightly higher excitation temperature of 10K. In addition, the improved angular resolution may also contribute to the higher beam-averaged C$^{18}$O column densities. The C$^{18}$O fractional abundances relative to H$_{2}$ are directly calculated by the ratio between C$^{18}$O and H$_{2}$ column densities. The latter is derived from the SED fitting with Herschel data (see Sect.~\ref{Sec:arc}). The distribution of the C$^{18}$O fractional abundances is shown in Fig.~\ref{Fig:abund}c. The C$^{18}$O fractional abundances range from 5$\times 10^{-8}$ to 3.5$\times 10^{-7}$ with a median of 1.6$\times 10^{-7}$. The map shows that the C$^{18}$O abundance drops toward the 1.1 mm dense cores. The region toward bolo3 and Ser-emb10 has the lowest abundance of $\sim$5$\times 10^{-8}$. This confirms the presence of C$^{18}$O depletion in this filament \citep{2018A&A...620A..62G}. Setting a C$^{18}$O fractional abundance of 2$\times 10^{-7}$ as a threshold for C$^{18}$O depletion in Fig.~\ref{Fig:abund}c, the areas (projected onto the plane of the sky) showing depletion are estimated to be about 0.5 pc$\times$0.12 pc in NW and 0.3$\times$0.1 pc in SE. NW has a larger depletion size than SE, which is likely because NW has more widespread dense gas. This reason also explains the fact that the C$^{18}$O depletion size in NW is larger than the typical depletion size ($\lesssim$0.1 pc) found in a single prestellar core \citep[e.g.,][]{2007ARA&A..45..339B}.

\begin{figure*}[!htbp]
\centering
\includegraphics[width = 0.95 \textwidth]{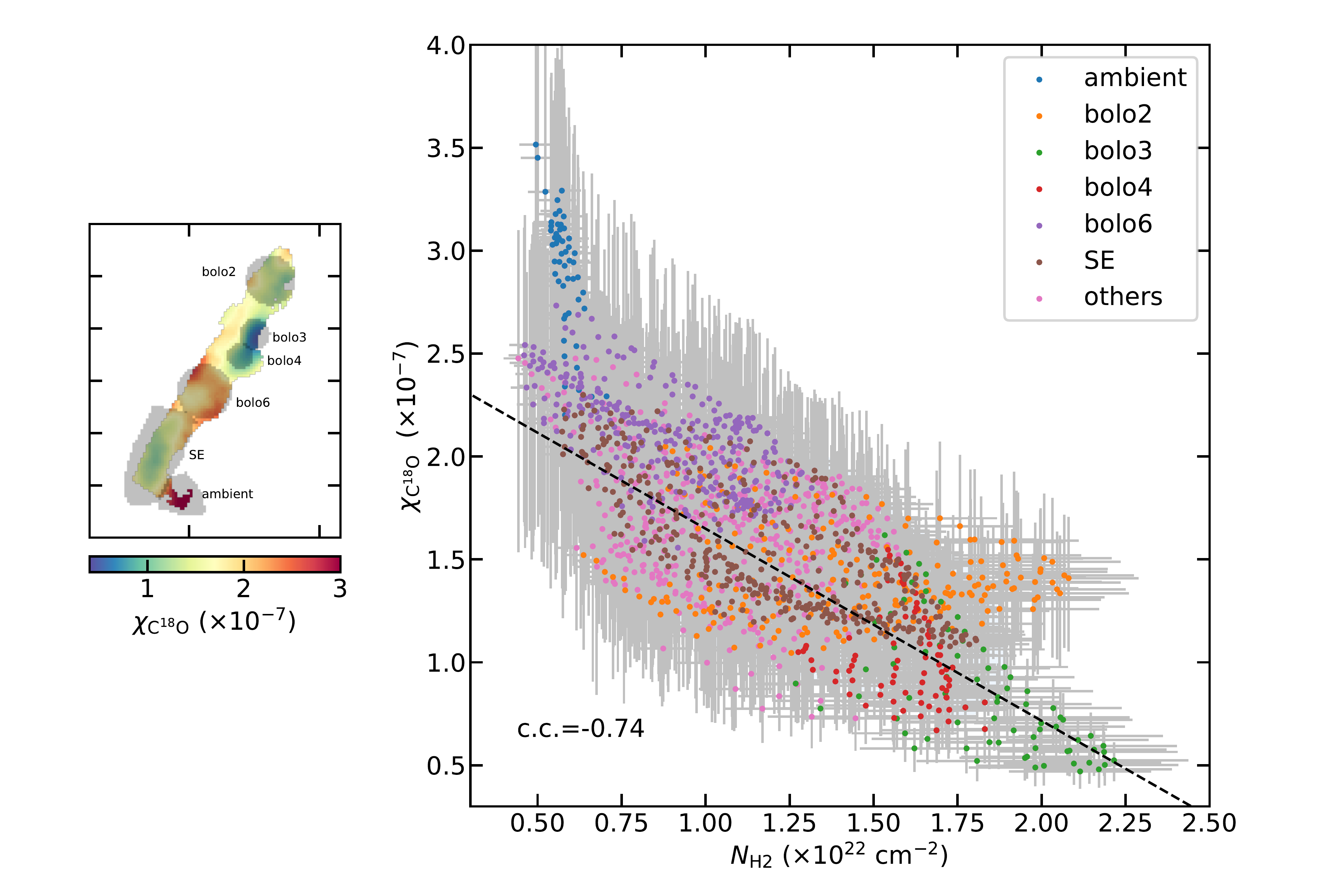}
\caption{{\textit{Left:} C$^{18}$O fractional abundance map with different labeled masks which correspond to the legend of the right panel. \textit{Right:} C$^{18}$O fractional abundance as a function of H$_{2}$ column density. The gray error bars indicate the measured uncertainties in C$^{18}$O fractional abundance and H$_{2}$ column density. 
The colors of the markers correspond to different regions indicated in the legend.
The Pearson correlation coefficient is shown in the lower left. 
The black dashed line represents the linear fit to the observed relation.}\label{Fig:col-abun}}
\end{figure*}

A pixel-by-pixel comparison between H$_{2}$ column densities and C$^{18}$O fractional abundances is displayed in Fig.~\ref{Fig:col-abun}. We find that the C$^{18}$O fractional abundance decreases with increasing H$_{2}$ column densities. This suggests that CO depletion onto dust grains becomes more efficient in denser regions, in agreement with theoretical predictions \citep[e.g.,][]{2007ARA&A..45..339B}. We performed a linear fit to obtain the empirical relation characterizing the decrease in the C$^{18}$O fractional abundances with increasing H$_{2}$ column densities, which gives $\frac{\chi_{\rm C^{18}O}}{1\times 10^{-7}} = (-0.93\pm 0.02)\frac{N({\rm H}_{2})}{1\times 10^{22} {\rm cm}^{-2}}+(2.67\pm 0.03)$. Furthermore, the lowest C$^{18}$O abundances of $\sim$5$\times 10^{-8}$ are about a factor of 6 lower than those ($\sim$3.0$\times 10^{-7}$) in the ambient gas. We also find that different regions show distinct distributions in Fig.~\ref{Fig:col-abun}. Intriguingly, there are two branches at H$_{2}$ column densities of $>$1.5$\times 10^{22}$ cm$^{-2}$. The upper branch arises from the bolo2 region, while the lower branch originates in the bolo3 region. Despite similar H$_{2}$ column densities, dust temperature, and C$^{18}$O rotational temperature, the bolo2 region shows a factor of $\gtrsim$2 higher C$^{18}$O fractional abundance than the bolo3 region. This might be caused by their different evolutionary stages with the bolo3 region being more evolved than the bolo2 region, (see discussion in Sect.~\ref{sec.model}). Alternatively, the H$_{2}$ number density of the bolo3 region might be enhanced by the outflow shock, resulting in more efficient CO depletion in the bolo3 region.


\section{Discussion}\label{Sec:dis}
\subsection{Local velocity gradients}\label{sec.vg}
Thanks to the improved angular resolution, we are able to investigate the local velocity gradients at a linear resolution of $\sim$0.07 pc. Because the velocity centroid maps agree well with each other (see Figs.~\ref{Fig:kine}--\ref{Fig:kine2}), we only use the highest signal-to-noise C$^{18}$O (2--1) data to derive the local velocity gradients. Following previous studies of dense cores \citep[e.g.,][]{1993ApJ...406..528G}, we determine the local velocity gradients in the Serpens filament by fitting the linear function:
\begin{equation}\label{f.grad}
  \varv_{\rm LSR} = \varv_{0} +a\Delta \alpha +b\Delta \delta\;,
\end{equation}  
where $\varv_{\rm LSR}$ is the observed LSR centroid velocity, $\varv_{0}$ is the systemic LSR velocity, $\Delta \alpha$ and $\Delta \delta$ are the offsets in right ascension and declination, and $a$ and $b$ are the components of the velocity gradient along the directions of right ascension and declination. The Levenberg-Marquardt algorithm was employed to fit this function toward each block with adjacent 3$\times$3 pixels where the LSR velocity has been successfully derived in order to calculate the local velocity gradient of the central pixel. 
The local velocity gradient $\nabla V$ has a magnitude of $|\nabla V| =\sqrt{a^{2}+b^{2}}$ and a position angle of $\theta_{\rm pa}={\rm arctan}(a/b)$. The position angle, $\theta_{\rm pa}$, increases counter-clockwise with respect to the north. The uncertainties in the local velocity gradients and their position angles are calculated through error propagation.

\begin{figure*}[!htbp]
  \centering
\includegraphics[width = 1.0 \textwidth]{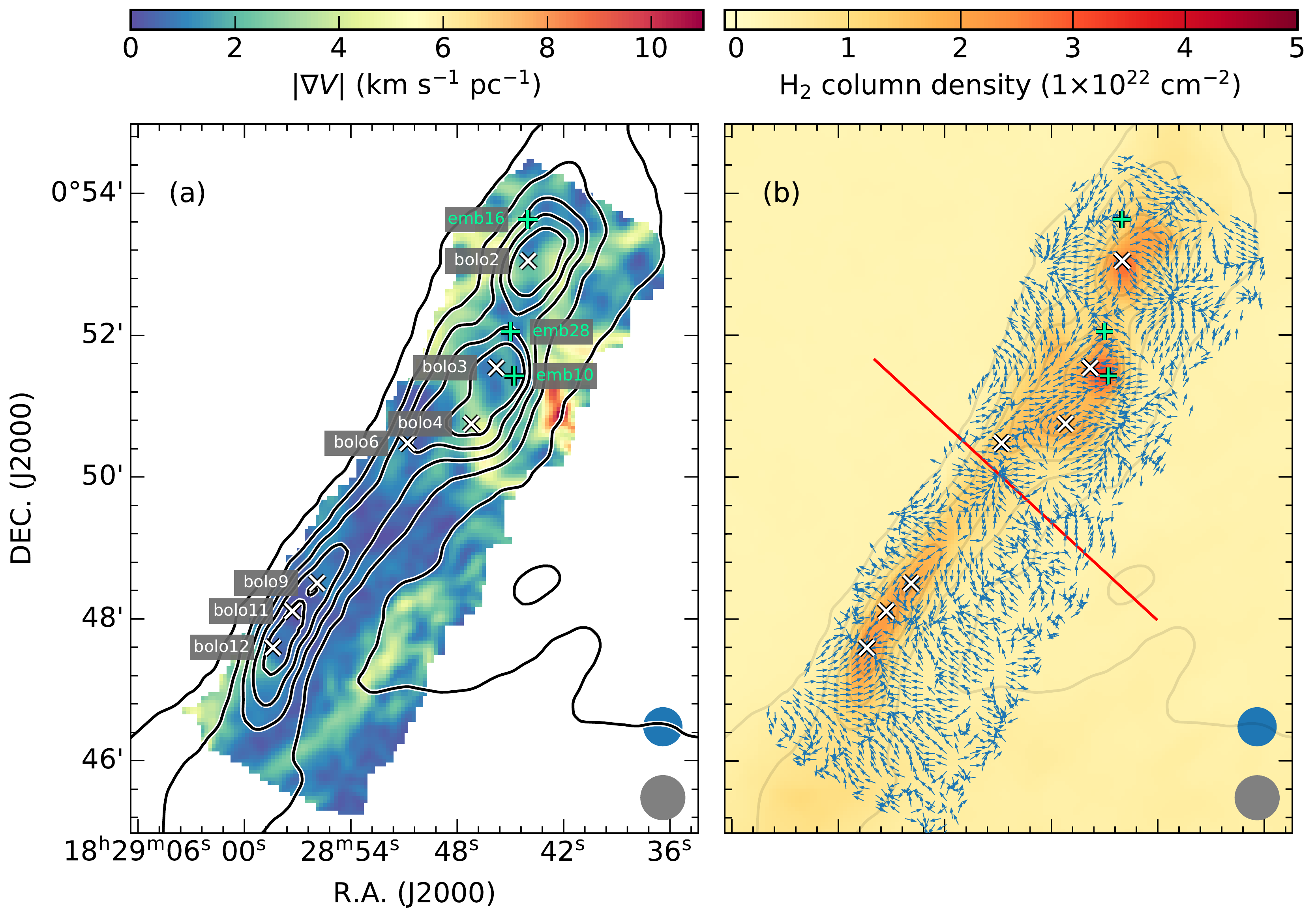}
\caption{{(a) Local velocity gradient magnitude map overlaid with the H$_{2}$ column density contours. (b) Herschel H$_{2}$ column density map overlaid with the normalized velocity vector maps. The arrows represent the estimated local velocity gradients which are rotated by 180\degree\,in order to better visualize the accretion directions in SE. The polarized angle of the Planck 353 GHz thermal dust emission has been rotated by 90\degree\,to trace the magnetic field direction which is indicated by the red line. In both panels, the contours correspond to H$_{2}$ column densities from 1$\times 10^{21}$~cm$^{-2}$ to 1.6$\times 10^{22}$~cm$^{-2}$ with a step of 3$\times 10^{21}$~cm$^{-2}$, and the beam sizes for the H$_{2}$ column density and velocity gradient maps are shown as the gray and blue circles in the lower right corner. The three green pluses give the positions of the three embedded YSOs, emb10, emb16, and Ser-emb 28 \citep{2009ApJ...692..973E}, and the white crosses mark the positions of the seven dust cores \citep{2007ApJ...666..982E}.}\label{Fig:vector}}
\end{figure*}

\begin{figure*}[!htbp]
\centering
\includegraphics[width = 0.95 \textwidth]{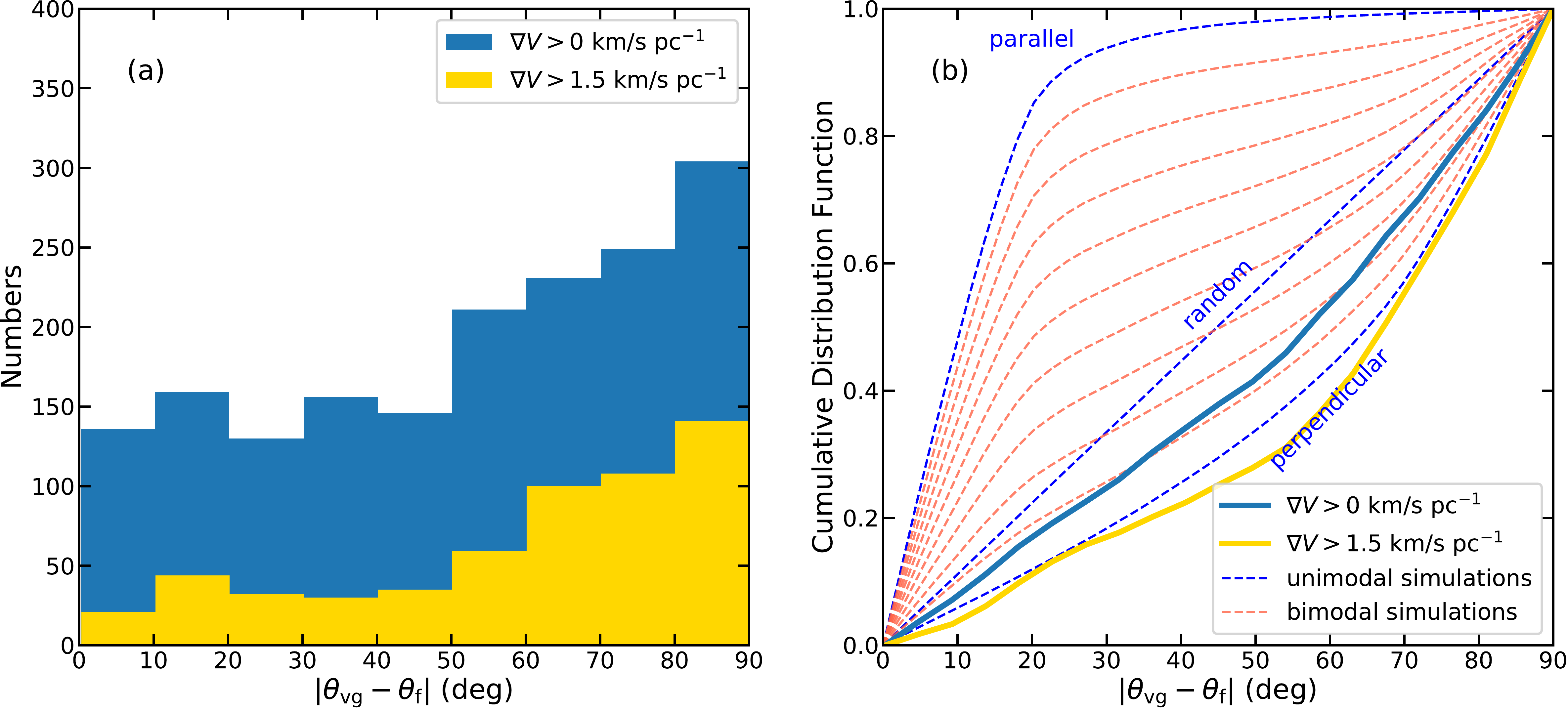}
\caption{{(a) Histograms of the angle of the vector of the observed local velocity gradient relative to the long axis of the filament. The blue histograms show all the angles of the observed velocity gradients in SE, while the yellow histograms represent the velocity gradients with $|\nabla V| >$1.5~\kms~pc$^{-1}$. (b) Cumulative distribution function of the angle between the velocity gradients and the major axis of the filament. The solid blue and yellow lines represent the observed cumulative distribution function of relative angles with corresponding and $|\nabla V| >$1.5~\kms~pc$^{-1}$ and $|\nabla V| >$1.5~\kms~pc$^{-1}$, respectively. The three blue dashed lines present the unimodal simulations of the expected angles which are three-dimensionally parallel (0--20$^{\circ}$), perpendicular (70--90$^{\circ}$), or random (0--90$^{\circ}$), while the red dashed lines show the bimodal simulations of the angles that are parallel and perpendicular with increments of 10\% (i.e., the top red line is 90\% parallel and 10\% perpendicular, the next one is 80\% parallel and 20\% perpendicular).}\label{Fig:ang}}
\end{figure*}

The derived $\nabla V$ distributions are shown in Fig.~\ref{Fig:vector}. The magnitude of $|\nabla V|$ varies from $\sim$0.1~\kms~pc$^{-1}$ around Bolo11 to $\sim$9~\kms~pc$^{-1}$ in the east of emb10. Most $|\nabla V|$ are higher than the values estimated from a large-scale analysis \citep[$\lesssim$1.5~\kms~pc$^{-1}$,][]{2018A&A...620A..62G}. Generally speaking, SE has a lower mean $|\nabla V|$ than NW. Because the velocity gradient can be regarded as the inverse of a timescale that it takes for the edge of the filament to fall into the center, the high $|\nabla V|$ may suggest that NW has higher accretion rates, leading to more active star-forming activities and higher turbulent motions as observed. This is also supported by the fact that higher $|\nabla V|$ are observed in active star-forming regions \citep{2014ApJ...797...76L,2019ApJ...886..119C,2020MNRAS.494.1971C}. Furthermore, it is evident that $|\nabla V|$ is generally higher in the filament's outskirts than its crest. NW shows higher $|\nabla V|$ and more chaotic $\theta_{\rm pa}$ than SE, which can be attributed to the YSO feedback and/or a greater impact of accretion onto the filament from ambient gas. Toward the more quiescent region SE, the local velocity gradient, $|\nabla V|$, decreases from the ambient cloud ($|\nabla V| >$2 \kms~pc$^{-1}$) to the crest ($|\nabla V| <$0.5 \kms~pc$^{-1}$). Figure~\ref{Fig:ang}a shows the statistical distribution of the angle between the filament's long axis \citep[$\theta_{\rm f}$=147$^{\circ}$,][]{2015A&A...584A.119R} and the local velocity gradient vectors. We obtained 47\% for angles ranging from 60\degree\,to 90\degree. If we only consider local velocity gradients with $|\nabla V| >$1.5 \kms~pc$^{-1}$, the percentage increases to 62\%. 

In order to test the observed trend, we carried out 3D Monte Carlo simulations which were then projected onto 2D to simulate the cumulative distribution function  (CDF) of the expected angles that are nearly parallel (0--20$^{\circ}$), nearly perpendicular (70--90$^{\circ}$), or completely random (0--90$^{\circ}$), following the method of unimodal and bimodal simulations introduced by previous studies \citep{2013ApJ...768..159H,2017ApJ...846...16S}. The bimodal distribution includes the expected angles that are both nearly parallel and nearly perpendicular. We considered 99 bimodal cases in steps of 1\%\, starting with 1\% parallel and 99\% perpendicular angles. Figure~\ref{Fig:ang}b compares the CDF of the observed angles and the simulated models. It is evident that the CDFs of the observed angles deviate from the unimodal model that is nearly parallel. The two CDFs of the observed angles are inconsistent with a purely parallel alignment at a $>$99\%\,confidence level with a corresponding p-value of $<$0.001 given by the Kolmogorov-Smirnov (KS) test. However, we cannot reject the null hypothesis for the other unimodal and bimodal simulations. For the observed CDFs with $\nabla V>$0 and $\nabla V >$1.5~\kms, the KS test gives a p-value of 0.82 and 0.93 against the unimodal model that is completely random, and 0.12 and 0.99 against the unimodal model that is nearly perpendicular. For the observed CDF with $\nabla V >$1.5~\kms, we also found high p-values of 0.90 against the bimodal simulations with $>$95\%\,perpendicular angles. This demonstrates that the velocity gradients with $\nabla V >$1.5~\kms are more likely perpendicular to the filament's long axis. As observed in other filaments \citep[e.g.,][]{2015A&A...584A..67B,2018ApJ...853..169D}, these vectors may be indicative of accreting molecular gas from its ambient cloud. 

The velocity gradients are indicative of motions of mass flows. Because material can be accreted from either blueshifted or redshifted velocities, the accretion flow is either in the same direction as the arrows representing the local velocity gradients in Fig.~\ref{Fig:vector}b (in the case of accretion from redshifted velocities, i.e., gas coming from the foreground), or in the opposite direction (in the case of accretion from blueshifted velocities, i.e., gas coming from the background). Inspecting Fig.~\ref{Fig:vector}b, we find that the vectors are clearly converging toward at least two dense cores, Bolo2 and Bolo12, and their magnitudes are changing significantly, ruling out the possibility of bulk motions being dominated by the filament's solid-body rotation. Furthermore, $\nabla V$ is also converging toward $\sim$30\arcsec\,south of Bolo6 where the H$_{2}$ column density is not enhanced. This indicates core accretion at a very early evolutionary stage. The $\nabla V$ vectors are also wrapping the filament's southern end, providing additional observational support for the edge effect of a finite filament \citep{2004ApJ...616..288B,2013ApJ...769..115H,2020A&A...637A..67Y}.

Because of the coarse angular resolution ($\sim$5\arcmin), the polarization of the Planck 353 GHz thermal dust emission is only employed to measure the mean magnetic field direction \citep{2015A&A...576A.105P,2015A&A...576A.104P}. The result is shown in Fig.~\ref{Fig:vector}b. It is evident that the mean magnetic field direction is perpendicular to the filament's long axis, and parallel to the velocity gradients of the mass flows accreted onto the filament's crest. Given the coarse angular resolution of the Planck polarization measurements, further higher angular resolution polarization observations are needed to better assess the role of magnetic fields in the Serpens filament.

\subsection{Longitudinal profiles}\label{sec.lon}
\begin{figure*}[!htbp]
\centering
\includegraphics[width = 1.0 \textwidth]{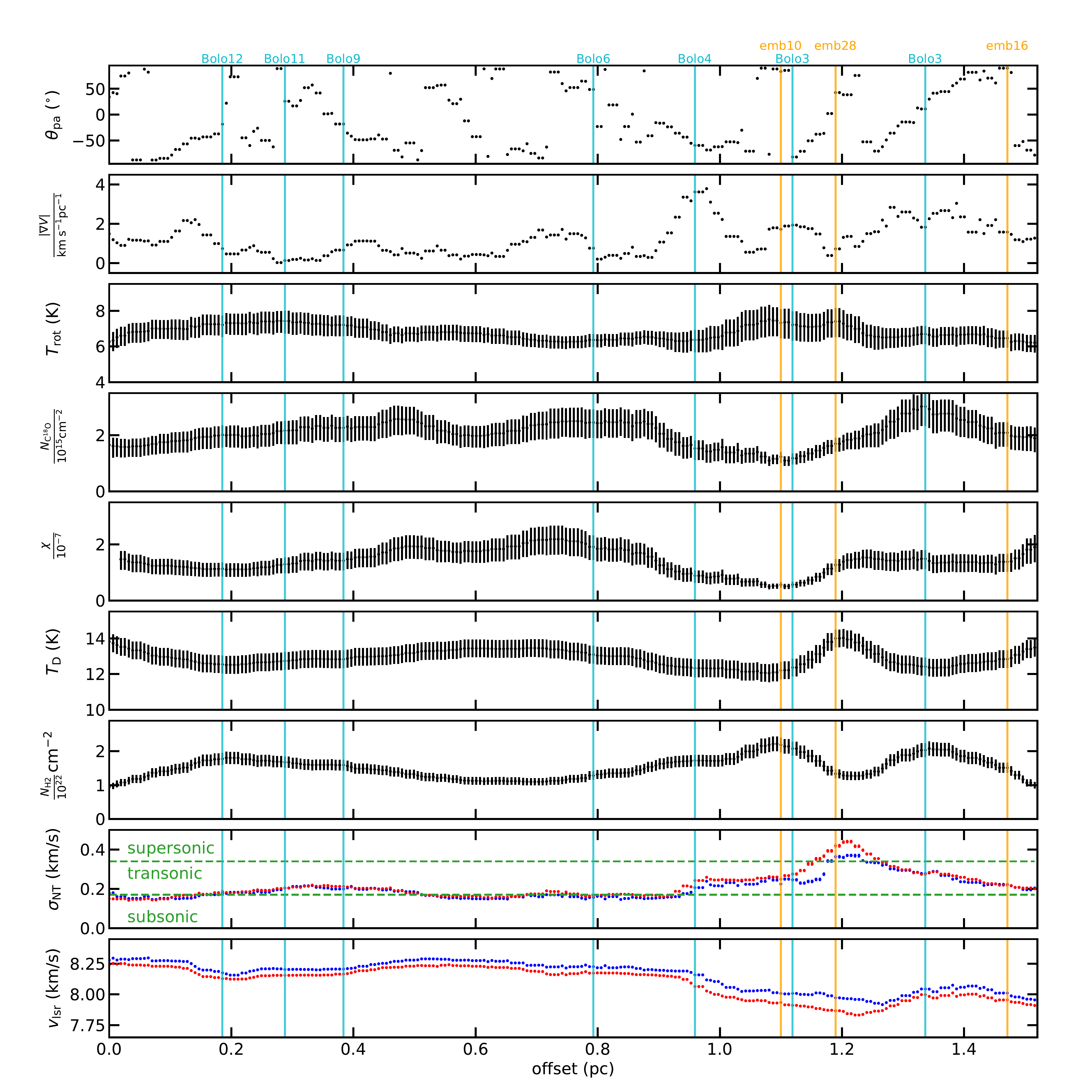}
\caption{{Physical and chemical parameters (from top to bottom: position angle of local velocity gradient, magnitude of local velocity gradient, C$^{18}$O rotational temperature, C$^{18}$O column density, C$^{18}$O fractional abundance, dust temperature, H$_{2}$ column density, nonthermal velocity dispersion, and velocity centroid) along the crest of the Serpens filament indicated by the blue line in Fig.~\ref{Fig:m0}d. The origin of the offsets corresponds to the southern end of the blue line in Fig.~\ref{Fig:m0}d. The offset increases from south to north. The positions which are close to the seven 1.1 mm dust cores and the three YSOs are indicated by the cyan and orange vertical lines, respectively. In the two lowest panels, the data derived from C$^{18}$O (1--0) and C$^{18}$O (2--1) are indicated by blue and red symbols, respectively.}\label{Fig:crest}}
\end{figure*}

Figure~\ref{Fig:crest} presents the C$^{18}$O rotational temperature, C$^{18}$O column density, C$^{18}$O fractional abundance, dust temperature, H$_{2}$ column density, nonthermal velocity dispersion, and velocity centroid\footnote{We find that generally the C$^{18}$O (1--0) line's LSR velocities are systematically $\approx 0.04$~\kms\, higher than the 
velocities of the C$^{18}$O (2--1) line. The uncertainties in the rest frequencies of the C$^{18}$O (1--0) and C$^{18}$O (2--1) lines are 6.3$\times 10^{-6}$ GHz and 1.5$\times 10^{-6}$ GHz, respectively; see Table \ref{Tab:lin}, whose values are taken from the CDMS \citep{2005JMoSt.742..215M}. This results in 1$\sigma$ uncertainties of 0.017~\kms\,and 0.002~\kms\,in velocity units. Hence, the systematic velocity difference may be caused (largely) by the fact the rest frequency we adopted for the 1--0 line is slightly too low. Given that the two lines were observed with different telescopes, there might also be a slight inaccuracy in the frequency calibration and calculation of one of the two telescopes.}
profile along the crest of the Serpens filament. We find that the rotational and dust temperatures have a small variation at offsets of $<$0.9 pc and their median values are 6.8 K and 12.9 K, respectively. This indicates that the crest is nearly isothermal with a kinetic temperature of $\sim$7 K, because the C$^{18}$O level populations are likely thermalized in the crest due to its high H$_{2}$ number density of $\sim$4$\times 10^{4}$~cm$^{-3}$ \citep{2018A&A...620A..62G}. The velocity centroids and nonthermal velocity dispersions are also nearly constant at offsets of $<$0.9 pc where the velocity gradients are $\lesssim$2~\kms~pc$^{-1}$. On the other hand, we neither see apparent velocity oscillation along the crest at a linear resolution of $\sim$0.07 pc, which were reported in other filaments \citep[e.g.,][]{2011A&A...533A..34H,2019MNRAS.487.1259L} nor velocity discontinuities expected in shock-induced structures \citep[e.g.,][]{2018MNRAS.479.1722C}. The observed nonthermal velocity dispersions become sonic at offsets of $<$0.9 pc when the sonic speed is assumed to be 0.17~\kms\,at a kinetic temperature of 7 K. This implies that most parts of the crest have a very low level of turbulence. These properties indicate that this filament can be approximately regarded as a finite isothermal and quiescent filament. 

Around the YSOs and 1.1 mm dust cores except Bolo4 in Fig.~\ref{Fig:crest}, we find a bimodal distribution of the position angle, $\theta_{\rm pa}$, of $\nabla V$. Around these targets, we find either a nearly flat $\theta_{\rm pa}$ distribution or a jump function. Depending on different geometries, accretion flows can come from either blueshifted or redshifted velocities from both sides, so such a bimodal distribution can be readily interpreted as a converging mass flow accreted by the YSOs or cores in an inhomogeous medium. The flat distribution may also be caused by cores' rotation. However, a recent study suggests that the velocity gradient at such a linear scale of 0.07 pc (14000 au) may not be dominated by rotation \citep{2020A&A...637A..92G}. On the other side, $|\nabla V|$ is low with a typical value of $<$2~\kms~pc$^{-1}$ around these YSOs and 1.1 mm dust cores except Bolo4. Bolo4 has the highest $|\nabla V|$ among these 1.1 mm dust cores, suggesting that Bolo4 has the highest accretion rate and will form the next protostar in this filament. Together with the fact that YSOs are located in its north and SE is more pristine, we speculate that star formation proceeds from north to south. Around Ser-emb28 (i.e., at offsets of $\sim$1.19 pc), we find elevated rotational temperature and dust temperature, but with lower H$_{2}$ column density. This is readily explained by the feedback of the outflow from Ser-emb28 (see Sect.~\ref{sec.outflow}). The outflow can sweep up the ambient molecular gas, resulting in a $N_{\rm H_{2}}$ drop at an offset of $\sim$1.2 pc and two enhanced $N_{\rm H_{2}}$ peaks at offsets of $\sim$1.1 pc and $\sim$1.35 pc. Furthermore, outflow shocks can heat the surrounding molecular gas, leading to elevated C$^{18}$O rotational and dust temperatures. One should note that the elevated dust temperature and reduced H$_{2}$ column density might alternatively arise from their degeneracy during the SED fitting. As mentioned above, C$^{18}$O molecules have been frozen out onto dust grains in most parts of the crest, but these C$^{18}$O molecules are sublimated or enhanced by sputtering of related species in shock regions, producing the observed higher C$^{18}$O column density and fractional abundance than in Ser-emb28's nearby siblings, emb10 and Bolo3. The high nonthermal velocity dispersions can also be explained by outflow shocks which drive additional turbulence (see Sect.~\ref{sec.outflow} and Appendix~\ref{append.a}). Furthermore, enhanced infall velocities may also contribute to the high nonthermal velocity dispersions. 


\subsection{Radial profiles}\label{sec.rad}
\begin{figure*}[!htbp]
\centering
\includegraphics[width = 0.8 \textwidth]{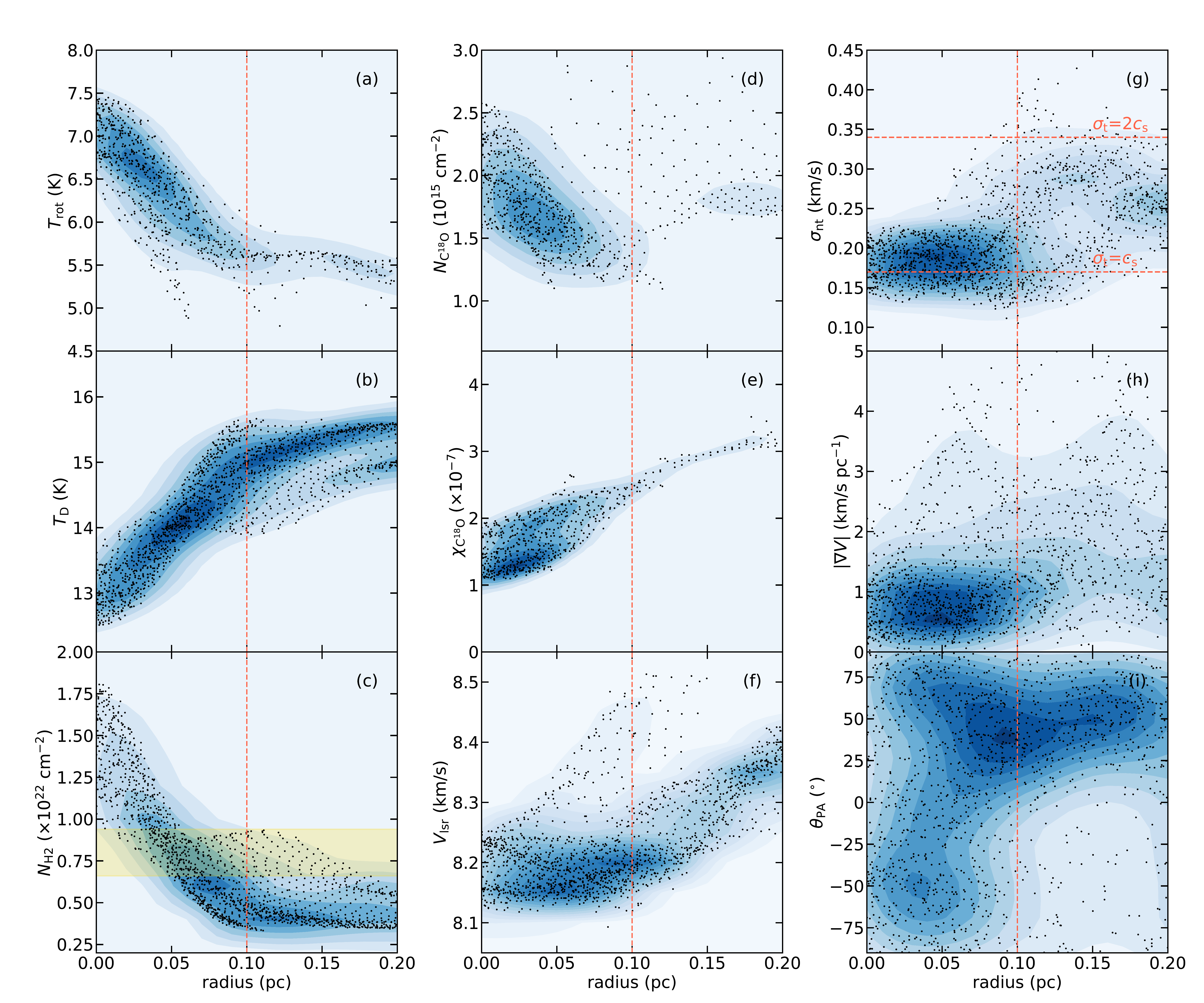}
\caption{{C$^{18}$O rotational temperature (Fig.~\ref{Fig:radial}a), dust temperature (Fig.~\ref{Fig:radial}b), H$_{2}$ column density (Fig.~\ref{Fig:radial}c), C$^{18}$O column density (Fig.~\ref{Fig:radial}d), C$^{18}$O fractional abundance (Fig.~\ref{Fig:radial}e), velocity centroid (Fig.~\ref{Fig:radial}f), nonthermal velocity dispersion (Fig.~\ref{Fig:radial}g), magnitude of velocity gradient (Fig.~\ref{Fig:radial}h), and position angle of velocity gradient (Fig.~\ref{Fig:radial}i) plotted as a function of their distance to the crest of SE. In Fig.~\ref{Fig:radial}c, the yellow shaded region marks the star formation threshold of $A_{\rm v}$=7--10 mag \citep[e.g.,][]{2004ApJ...611L..45J,2010ApJ...724..687L,2010ApJ...723.1019H,2010A&A...518L.102A,2015A&A...584A..91K}. In all panels, the radius of 0.1 pc is marked with a red dashed line, which indicates the transition discussed in Sect.~\ref{sec.rad}. We note that the data points become incomplete at larger radial distances in Figs.~\ref{Fig:radial}a, \ref{Fig:radial}d, \ref{Fig:radial}e, and\ref{Fig:radial}f due to the 3$\sigma$ clipping mentioned in Sect.~\ref{sec.abund}. In Fig.~\ref{Fig:radial}g, the two red horizontal dashed lines mark $\sigma_{\rm t}=c_{\rm s}$ and $\sigma_{\rm t}=2c_{\rm s}$ where $c_{\rm s}$ is the sound speed at a kinetic temperature of 7 K.}\label{Fig:radial}}
\end{figure*}

In order to avoid the influence of stellar feedback from YSOs in NW, we only investigate the radial distribution toward SE that is more pristine than NW. Figure~\ref{Fig:radial} presents the radial dependencies of the different physical and chemical properties. Figure~\ref{Fig:radial}a suggests that the rotational temperature profile increases monotonically toward the crest, which is caused by the variation of the H$_{2}$ number density. The observed rotational temperature profile validates the assumption used to interpret the observed blue-skewed profiles in filaments as the result of infall motions \citep[e.g.,][]{2013ApJ...766..115K,2018A&A...620A..62G}. Unlike the rotational temperature profile, the dust temperature profile decreases monotonically toward the spine (see Fig.~\ref{Fig:radial}b), similar to the results reported in other studies \citep[e.g.,][]{2011A&A...529L...6A}. This can be because the contribution of the foreground and background warm dust emission decreases toward the spine.

In Fig.~\ref{Fig:radial}c, we also find that the radius of $r\sim$0.05 pc corresponds to the star formation threshold of $A_{\rm v}$=7--10 mag, above which star formation can only occur \citep[e.g.,][]{2004ApJ...611L..45J,2010ApJ...724..687L,2010ApJ...723.1019H,2010A&A...518L.102A,2015A&A...584A..91K}\footnote{We also note that the existence of the star formation threshold is still under debate \citep{2018A&A...619A..52B}.}. Since the velocity dispersions drop to nearly sonic levels (see Fig.~\ref{Fig:radial}g) at this radius, this supports the scenario that the transition to the sonic turbulence can also be the reason for the origin of the star formation threshold. Figure~\ref{Fig:radial}d shows an increasing $N_{\rm C^{18}O}$ toward the crest while Fig.~\ref{Fig:radial}e shows a decreasing $\chi_{\rm C^{18}O}$ toward the crest. The former one is expected with increasing H$_{2}$ column densities, while the latter one can be explained by CO depletion that becomes higher in denser regions. 

Figure~\ref{Fig:radial}f presents the radial dependence of velocity centroids. The outer part is redshifted with respect to the inner part, pointing out that mass flows accrete material predominantly from the foreground gas. We can also infer that the velocity gradients are higher in the outer part. Furthermore, the rather smooth distribution also implies that there are no strong shocks. The radial velocity difference can be used to assess the motions in the filament. As introduced by \citet{2020MNRAS.494.3675C}, the ratio between the kinetic energy of the flow transverse to the filament and the gravitational potential energy of the filament gas is:
\begin{equation}\label{f.cv}
C_{\varv} = \frac{\Delta \varv_{\rm h}^{2}}{GM(r)/L}\,,
\end{equation}
where $\Delta \varv_{\rm h}$ is half the velocity difference across the filament out to a radius r, and $G$ is the gravitational constant, and $M(r)/L$ is the mass per unit length of the filament.  They pointed out that the local turbulence is more important than self-gravity when $C_{\varv} \gg 1$ while the velocity structure is likely induced by self-gravity when $C_{\varv} \lesssim 1$. Based on Fig.~\ref{Fig:radial}f, we find $\Delta \varv_{\rm h}\sim$0.17~\kms\,for $r\sim$0.17 pc where $M(r)/L$ was found to be 45~$M_{\odot}$~pc$^{-1}$. We thus obtain $C_{\varv}\sim$0.15, much lower than unity. Hence, this supports gravity-driven motions in the filament.

Figure~\ref{Fig:radial}g presents the radial dependence of the nonthermal velocity dispersions. We find a characteristic radius of $\sim$0.1 pc. The nonthermal velocity dispersion is nearly sonic at $r<$0.1 pc, while the nonthermal velocity dispersion can have higher values and becomes more scattered at larger radii. The observed low turbulence in the inner region surrounded by more turbulent outer layers can be readily explained by the fact that turbulence has higher energy dissipation rates in denser regions \citep[e.g.,][]{2004ARA&A..42..211E,2017MNRAS.465..667L}. This also supports the sonic scale of turbulence as the origin of the universal width of $\sim$0.1 pc found in nearby molecular filaments \citep{2011A&A...529L...6A,2019A&A...621A..42A}. On the other hand, the nonthermal velocity dispersion is nearly constant at $<$0.1 pc, which can be sustained by an internally driven turbulence, for instance, accretion-driven turbulence \citep{2020MNRAS.495..758H,2020ApJ...890..157X}.  On the other hand, our result is opposite to the result in the massive filament DR21 which shows increasing velocity dispersion toward the crest \citep{2010A&A...520A..49S}. This may be due to the different dynamical state of the DR21 filament or limited angular resolutions which are unable to resolve (sub)sonic structures.  

Figure~\ref{Fig:radial}h presents the radial profile of $|\nabla V|$ that is decreasing toward the spine. Such a radial dependence of $|\nabla V|$ was also reported in the velocity-coherent filaments in NGC 1333 \citep{2020ApJ...891...84C}. They proposed that the radial dependence of $|\nabla V|$ is caused by the ongoing accretion damped by the high density material as the accreting gas moves closer to the filament spine. Alternatively, this can be a geometric effect (more discussions are presented in Sect.~\ref{sec.formation}). Fig.~\ref{Fig:radial}i shows that the position angles of $\nabla V$ become ordered at a distance of $r\sim$0.1 pc away from the spine while they are more randomly distributed in the inner part at $r<$0.1 pc. This is also evident in Fig.~\ref{Fig:vector}b. Since the uncertainties of the position angles are even smaller in the inner part than in the outskirts, the position angle transition is thus an intrinsic result of accretion flows. The ordered $\nabla V$ can be channeled by magnetic fields at $r\sim$0.1 pc, while magnetic fields may become less important in channeling accretion flows at $r<$0.1 pc (for a more detailed discussion, see Sect.~\ref{sec.formation}). 

\subsection{Widespread C$^{18}$O depletion and its implications}\label{sec.model}
\subsubsection{Chemical model}
As mentioned in Sect.~\ref{sec.abund}, our observations confirm the widespread C$^{18}$O depletion, and also reveal a trend that the C$^{18}$O fractional abundances decrease with increasing H$_{2}$ column densities. This can be attributed to the H$_{2}$ density dependence of the CO freeze-out timescale. In order to simulate the observed trend, we run a set of models using the pseudo-time dependent chemical model, Chempl \citep{2020arXiv200711294D}. The 2012 version of the UMIST network \citep{2013A&A...550A..36M} is used, augmented by adsorption, desorption, and grain surface reactions. The initial conditions are assumed to be the standard oxygen-rich initial elemental abundances with every element in atomic form except that hydrogen is in H$_{2}$ \citep[see Table 3 of][]{2013A&A...550A..36M}. The interstellar ultra-violet radiation field is set to be the standard value in the solar neighborhood \citep{1978ApJS...36..595D} and the cosmic-ray ionization rate is assumed to be the canonical value of 1.36$\times 10^{-17}$ s$^{-1}$ \citep[e.g.,][]{2000A&A...358L..79V}. The self-shielding of H$_{2}$ is implemented using the analytical formula of \citet{1996ApJ...468..269D}, and the self-shielding of CO is treated with the ``one-band'' approximation \citep{1983ApJ...264..546M}. Based on the rotational temperature of the thermalized C$^{18}$O level populations (see Fig.~\ref{Fig:crest}), the kinetic temperature is simply assumed to be 7 K as the fiducial case. We adopted an isotopic ratio [$^{16}$O/$^{18}$O] of 530 \citep{1994ARA&A..32..191W,2012A&A...544L..15B} to convert the $^{12}$CO fractional abundance to the C$^{18}$O fractional abundance. Only two parameters, visual extinction ($A_{\rm v}$) and H$_{2}$ number density, are free parameters. Since the filament is likely formed from a sheet-like molecular cloud (see Sect.~\ref{sec.formation}), we thus assumed a constant thickness, and the value is set to be 0.68 pc, four times the filament's FWHM width derived from the dust-based H$_{2}$ column density profile \citep{2018A&A...620A..62G}. The H$_{2}$ number density can be determined by dividing the H$_{2}$ column density by the thickness. We ran 12 models with $A_{v}$ ranging from 2.12 to 25.4 mag. We adopt the extinction relationship as $N({\rm H_{2}})$ = $9.4\times 10^{20}$ ($A_{\rm v}$/mag) \citep{1978ApJ...224..132B}.

\begin{figure*}[!htbp]
\centering
\includegraphics[width = 0.95 \textwidth]{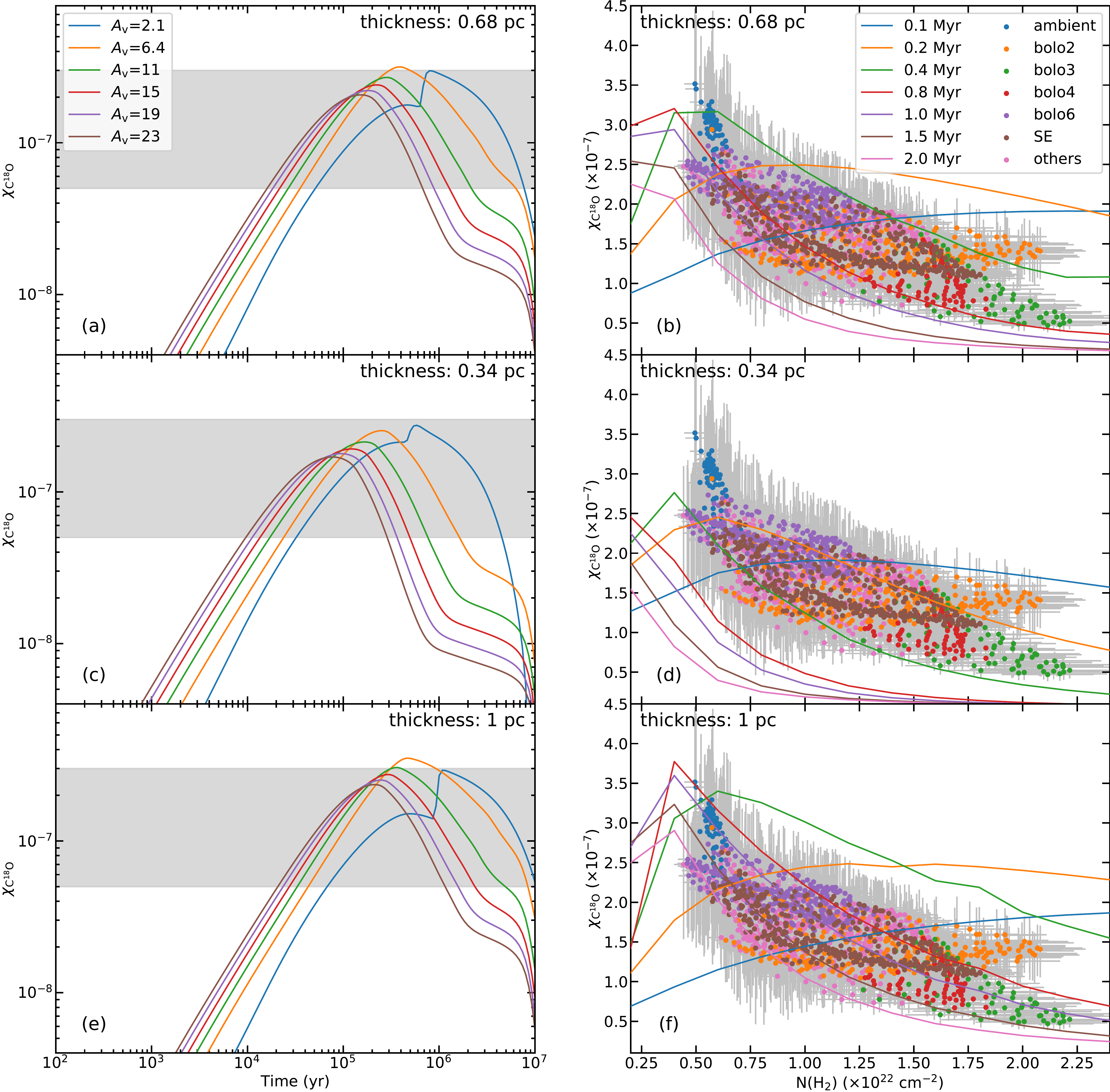}
\caption{{(a) C$^{18}$O fractional abundances, $\chi_{\rm C^{18}O}$, as a function of time, as calculated with different H$_{2}$ column densities and an assumed thickness of 0.68~pc. The shaded region corresponds to the observed C$^{18}$O fractional abundance range. (b) Similar to Fig.~\ref{Fig:col-abun}, but overlaid with the isochronic lines which are based on the results in Fig.~\ref{Fig:chemt}a. The markers represent the observed relationship between C$^{18}$O fractional abundance and H$_{2}$ column density with different regions labeled in different colors. Figures~(c) and (e) are similar to Fig.~\ref{Fig:chemt}a, but for an assumed thickness of 0.34~pc and 1 pc, respectively. Figures~(d) and (f) are similar to Fig.~\ref{Fig:chemt}b, but for an assumed thickness of 0.34~pc and 1 pc, respectively.}\label{Fig:chemt}}
\end{figure*}

The pseudo-time dependent chemical modeling results are shown in Fig.~\ref{Fig:chemt}a. This demonstrates that higher H$_{2}$ column density regions tend to be depleted earlier, as expected. Based on the modeling results, we utilize the isochronic tracks in the $N$(H$_{2}$)--$\chi_{\rm c^{18}o}$ plot (see Fig.~\ref{Fig:chemt}b) in order to constrain the chemical timescale of the Serpens filament. The two branches from the bolo2 and bolo3 regions (see Sect.~\ref{sec.abund}) are characterized by $\sim$0.4 Myr and $\sim$0.8 Myr, respectively, suggesting that the bolo3 region is more evolved than the bolo2 region. Overall, the observed values lie within the chemical timescale of $\lesssim$1.5 Myr. Intriguingly, the trend of the observed distribution is roughly reproduced by the isochronic track of $\sim$0.8 Myr. However, we also note that the assumed thickness is weakly constrained and can lead to a large uncertainty in the chemical timescale. The assumed thickness directly implies the H$_{2}$ number density used in the models. If we adopt a lower value for thickness, the H$_{2}$ number density will become higher, leading to a shorter timescale. On the contrary, a greater thickness will give rise to a longer timescale. In order to investigate this impact, we run the models with a smaller thickness of 0.34 pc, twice the filament's FWHM width, and a greater thickness of 1 pc. The results are shown in Figs.~\ref{Fig:chemt}c--\ref{Fig:chemt}f. This demonstrates that the CO depletion is dependent on the assumed thicknesses, that is, a greater thickness will result in a longer timescale. On the other hand, the high C$^{18}$O fractional abundances of $\sim$3$\times 10^{-7}$ cannot be reproduced by the model using a thickness of 0.34 pc, and a greater thickness is more plausible for these regions. For the thickness of 1 pc, the chemical timescale increases to $\lesssim$2 Myr. Because the freeze-out timescale is weakly dependent on gas temperature \citep[e.g.,][]{2013A&A...550A..36M} and  the reaction rates would not change significantly with a small change in gas temperature, effects of  temperature variation are neglected in our case. We also neglect the inhomogeneity of the molecular cloud which can give rise to the scatter in the $N$(H$_{2}$)--$\chi_{\rm c^{18}o}$ plot. We have to note that the density evolution is not taken into account in our models, which could prolong the chemical timescale. H$_{2}$ formation is a prerequisite to CO formation in chemical models \citep[e.g.,][]{2007ARA&A..45..339B}, and the chemical timescales do not include any earlier phase when gas is predominantly molecular (e.g., H$_{2}$) but CO is not abundant enough for detectable emission.


\subsubsection{Evolutionary stage}\label{sec.timescale}
Different accretion models can also allow us to estimate and constrain evolutionary timescales. If we assume that this filament starts accreting mass from its surrounding material with an initial thermal critical line mass of 16~$M_{\odot}$~pc$^{-1}$, the characteristic accretion timescale to reach the observed line mass of 36--41 $M_{\odot}$~pc$^{-1}$ is on the order of 0.5--1 Myr based on the time-dependent accretion model \citep[see Appendix C in][]{2019A&A...629L...4A}. For a constant accretion model, the timescale can be as low as about 0.3 Myr if we assume a constant infall rate of 72~$M_{\odot}$~Myr$^{-1}$ \citep{2018A&A...620A..62G}. Because the initial gas in the chemical models is nearly atomic, the chemical models include processes prior to the accretion phase mentioned above. Therefore, these timescales are roughly consistent with our derived chemical timescale of $\lesssim$2 Myr. On the other hand, the presence of Class I YSOs indicates an age of about 1--1.5 Myr, because statistical studies suggest that the lifetime of prestellar and Class 0+I phases are about 1 Myr and 0.5 Myr \citep{2014prpl.conf..195D,2015A&A...584A..91K}, respectively. This age is comparable to our chemical timescale. 

The fact that the accretion timescale coincides with the chemical timescale provides hints concerning the origin of the Serpens filament. First, we note that these timescales do not necessarily need to agree with each other. For example, for a non-accreting filament, its accretion timescale would reach infinity and would far exceed the chemical timescale. In our case, both timescales seem to be comparable, which means that the accretion is occurring and contributing to its mass growth significantly. This further suggests that the Serpens filament, although only slightly supercritical, is still growing and might end up as a much more massive filament before being dispersed. 






\subsection{Filament formation and evolution}\label{sec.formation}
The Serpens filament is a long velocity-coherent (tran)sonic filament, showing similar properties to fibers. Many models have been proposed to explain such (tran)sonic filaments in molecular clouds. In the ``fray and gather" model \citep{2014MNRAS.445.2900S,2016MNRAS.455.3640S}, filaments form as a consequence of the turbulent cascade in molecular clouds. The simulations of \citet{2016MNRAS.455.3640S} suggest that fibers form first. As the large-scale turbulence decays, gravity becomes more dominant, gathering the adjacent fibers to form a main filament. In the ``fray and fragment" model \citep{2015A&A...574A.104T,2017MNRAS.468.2489C}, a filament forms by the supersonic collision of two flows, and ``frayed" to produce intertwined subsonic fibers. Gravitationally unstable fibers further fragment into dense cores that collapse to form nascent stars. Filaments can be also created by a magneto-hydrodynamic (MHD) shock wave in an inhomogeneous cloud \citep{2013ApJ...774L..31I,2018PASJ...70S..53I,2018PASJ...70...96A,2020A&A...644A..27B}. We note that the MHD shock wave is usually assumed to be created by colliding flows, similar to the ``fray and fragment" model. Furthermore, a dense filament can form in a shell created by a large-scale compression \citep[e.g.,][]{2019A&A...623A..16S}. Although there is no indication for substructures parallel to the main filament in the Serpens filament, this could result from the limited angular resolutions of our observations. Therefore, we are not able to exclude or discriminate between the ``fray and gather" and ``fray and fragment" models. On the other hand, we did not find any indication of strong shocks which are expected in the MHD shock scenario. However, this could be because the shock-induced energies have been already dissipated at an earlier stage, leading to the observed (tran)sonic motions. We also note that MHD shocks tend to create filaments perpendicular to the magnetic field \citep[see Fig. 3 in][]{2018PASJ...70S..53I}, as indicated by the observed large-scale polarization direction (see Fig.~\ref{Fig:vector}b). In our observations, we do not find shell structures at the current linear resolution ($\sim$0.07 pc), so the large-scale compression is not a likely scenario for the formation of the Serpens filament. Therefore, we propose that turbulence or colliding flows may best explain the filament formation.

\begin{figure*}[!htbp]
\centering
\includegraphics[width = 0.9 \textwidth]{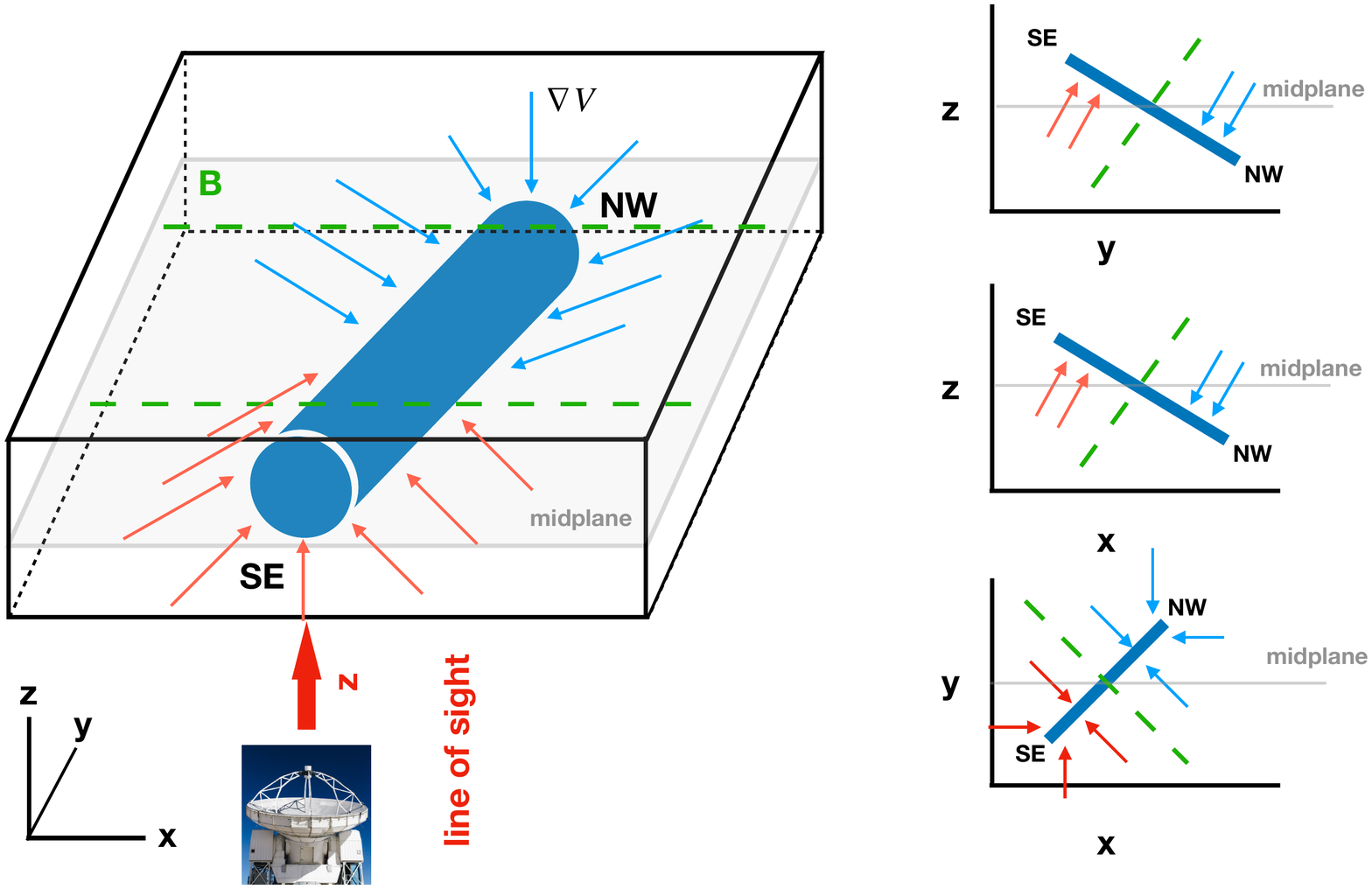}
\caption{{Schematic view of the geometry of the Serpens filament. The left panel gives a perspective view of its 3D geometry, and the right panels represent the projection on to the x-y, x-z, and y-z planes. The midplane and its parallel surfaces have the same distances to observers, and the line of sight is along the z direction. The local velocity gradients that are associated with blueshifted and redshifted flows are indicated by the blue and red arrows, while the magnetic field is indicated by green dashed lines.}\label{Fig:model}}
\end{figure*}

Sheet-like structures are very common in numerical simulations, and they can be easily created by shocks \citep[e.g.,][]{2006ApJ...643..245V,2020MNRAS.497.4196S}. Previous observations also indicate that molecular clouds can be sheet-like \citep[e.g.,][]{2015ApJ...811...71Q,2019A&A...623A..16S}. Thus, we  assume that the filament is formed in a sheet-like molecular cloud. In order to explain the observed local velocity gradients, we hypothesize that the filament is formed not in the midplane of the sheet, but inclined with SE at the far side and NW at the near side with respect to the observer, as shown in Fig.~\ref{Fig:model}. The proposed turbulence or colliding flows in an inhomogeneous cloud could  create a filament in such a geometry. Once the initial filament is formed, accretion flows can be driven onto the filament due to its greater gravitational potential. Thus, our geometry hypothesis could readily explain the observed redshifted ambient gas in SE and the blueshifted ambient gas in NW. The momentum of these flows would make the velocity difference between the filament and ambient gas greater with evolution. The current velocity difference is only about 0.5~\kms, implying an early evolutionary stage.


As shown in Fig.~\ref{Fig:vector}b, the observed $\nabla V$ in the outskirts is parallel to the mean magnetic field that is perpendicular to the filament's long axis. In nearly magnetically critical regions, molecular gas tends to move along magnetic field lines, because the Lorentz force tends to resist gas flows in the perpendicular direction \citep{2008ApJ...687..354N,2014prpl.conf..101L}. In case that the filament is close to magnetically critical conditions, we can provide a rough estimate of the magnetic strength. Following Eqs.~(I--II) in \citet{2014prpl.conf..101L}, we obtain a magnetic strength of about 8--16~$\mu$G in the corresponding regions with an H$_{2}$ column density of $\sim$4$\times 10^{21}$~cm$^{-2}$. In the inner region closer to the crest, molecular gas becomes magnetically supercritical, that is, gravity dominates over the magnetic fields. This leads to the observed converging $\nabla V$ morphology around Bolo2 and Bolo12. This case can also explain the observed transition that the position angle of $\nabla V$ becomes more random at $r\lesssim 0.1$ pc (see Fig.~\ref{Fig:radial}i). Alternatively, simulations show that the accretion flows are able to drag the magnetic field lines along the flow in the magnetically supercritical regime, causing it to be oriented perpendicular to the filaments \citep{2018MNRAS.480.2939G}. In this case, the magnetic strength should be lower than 8--16~$\mu$G. 



The radial dependence of $\varv_{\rm lsr}$ and $|\nabla V|$ observed in Figs.~\ref{Fig:radial}f and \ref{Fig:radial}h can be at least partially a result of geometric effects. In a symmetric cylinder, one would expect a nearly constant $\varv_{\rm lsr}$ which is the density-weighted 3D velocity averaged over the line of sight. At $r<$0.1 pc, the geometry is dominated by such a symmetric cylinder, leading to the nearly constant $v_{\rm lsr}$ and low $|\nabla V|$. At $r>$0.1 pc, accretion from ambient gas becomes more important. Hence, more redshifted $v_{\rm lsr}$ and higher $\nabla V$ are observed at a greater radial distance.

The Serpens filament is apparently truncated. Different from infinitely-long filaments, material tends to pile up at finite filaments' ends due to the edge effect, also known as gravitational edge focusing \citep{2004ApJ...616..288B,2013ApJ...769..115H}. Because of the accumulated mass, the southern end becomes magnetically supercritical. Hence, the observed $\nabla V$ is wrapping the southern end of this filament (see Fig.~\ref{Fig:vector}b), rather than parallel to the observed mean magnetic field. Furthermore, the edge effect also gives rise to the steep component in Figs.~\ref{Fig:radial}f and \ref{Fig:radial}h because of greater gravitational potentials.

\section{Summary and conclusion}\label{Sec:sum}
We have performed $^{13}$CO (1--0), C$^{18}$O (1--0), C$^{17}$O (1--0), $^{13}$CO (2--1), C$^{18}$O (2--1), and C$^{17}$O (2--1) line imaging toward the Serpens filament with the IRAM-30 m and APEX-12 m telescopes in order to characterize its physical and chemical structure with a linear resolution of $\sim$0.07 pc and a spectral resolution of $\lesssim$0.1~\kms. Our observations have led to the following main results:
\begin{itemize}

\item[1.] We reveal widespread $^{13}$CO (2--1) self-absorption, resulting in a markedly different morphology relative to the filamentary structure traced by C$^{18}$O and C$^{17}$O. We discover a new molecular outflow driven by the Class I protostar Ser-emb28 in $^{13}$CO (2--1). This outflow is likely to exert feedback on the filament, driving a small amount ($\sim$5\%) of turbulent energy in this filament. The energy injection rate from the outflow is comparable to the filament's energy dissipation rate, indicating that the outflow can sustain the observed turbulence.

\item[2.] Averaging C$^{18}$O and C$^{17}$O spectra in the low-brightness temperature regions, we derive the isotopic ratio [$^{18}$O/$^{17}$O] to be 3.94$\pm$0.23 and $4.41\pm$0.24 for $J=$1--0 and $J$=2--1, respectively. Assuming local thermodynamic equilibrium and fixing the isotopic ratio [$^{18}$O/$^{17}$O] to be 4.1, we constrain the opacity, rotational temperature, and molecular column density for C$^{18}$O and C$^{17}$O simultaneously with Monte Carlo Markov chain calculations. We find that the opacities of C$^{18}$O (1--0) and C$^{18}$O (2--1) become higher than unity in most regions. The rotational temperature is determined to be 4.6--7.6~K, while the C$^{18}$O column densities range from 9.1$\times 10^{14}$ cm$^{-2}$ to 3.0$\times 10^{15}$ cm$^{-2}$. The C$^{18}$O fractional abundances relative to H$_{2}$ range from 5.8$\times 10^{-8}$ to 3.5$\times 10^{-7}$, confirming CO depletion in this filament. 

\item[3.] Based on the velocity centroid map, we derive local velocity gradients toward this filament. Velocity gradients decrease toward the filament's crest while the position angles are randomly distributed at a radius of $<$0.1 pc and become ordered at a radius of $\sim$0.1 pc. We performed Monte Carlo simulations to confirm that the local velocity gradients have a tendency to be perpendicular to the filament’s long axis in the outskirts. Velocity gradient vectors with relatively high magnitudes are wrapping the filament's southern end, providing additional observational support for the edge effect of a finite filament. 

\item[4.] We investigate the longitudinal and radial profile of physical and chemical parameters of the Serpens filament. The southeastern part of the filament can be regarded approximately as a finite isothermal and quiescent filament, while its northwestern part is likely affected by the outflow feedback of Ser-emb28. Unlike the dust temperature, the C$^{18}$O rotational temperature increases toward the filament's crest. At a radius of $\lesssim$0.1 pc, the nonthermal velocity dispersion is nearly sonic, less turbulent than the region farther out. The local velocity gradients are lower and more randomly-oriented at $r\lesssim 0.1$ pc than at $r> 0.1$ pc.

\item[5.] We utilize a pseudo-time dependent chemical model in order to model the CO depletion and constrain the chemical timescale of the Serpens filament. We obtain a timescale of $\lesssim$2 Myr, which is roughly comparable to the accretion timescale. This suggests that the filament is young and still accreting. The isochronic evolutionary track of the C$^{18}$O freeze-out process may serve as an independent tool to evaluate the evolution of filaments. 
\end{itemize}

On the basis of these results, we propose that the Serpens filament is a finite filament formed in a sheet-like molecular cloud, with a filamentary density enhancement first generated by turbulence or colliding flows, and then followed by gravity-driven accretion of ambient gas. Higher angular resolution observations of magnetic fields will be needed to further assess the role of the magnetic field which may affect the observed gravity-driven accretion flows.
  


\section*{ACKNOWLEDGMENTS}\label{sec.ack}
We wish to thank the IRAM and APEX staff for their assistance with our observations. This work was supported by the National Key R\&D Program of China under grant 2017YFA0402702. The research leading to these results has received funding from the European Union’s Horizon 2020 research and innovation program under grant agreement No 730562 [RadioNet]. This work is based on observations carried out under project numbers 102-17 and 150-18 with the IRAM 30m telescope. IRAM is supported by INSU/CNRS (France), MPG (Germany) and IGN (Spain). This publication is based on data acquired with the Atacama Pathfinder Experiment (APEX). APEX is a collaboration between the Max-Planck-Institut fur Radioastronomie, the European Southern Observatory, and the Onsala Space Observatory. Y.G. thanks Henrik Beuther, Philippe Andr{\'e}, and Alvaro Hacar for helpful discussions during the APEX2020 meeting. Y.G. thanks Michael Rugel and Aiyuan Yang for organizing the treasure hunt in Valle de la Luna during the APEX observations. Y.G. also thanks his wife, Wenjin Yang, for her suggestion to use the color style of Piet Mondrian in this paper. F.~Du is financially supported by the Hundred Talents Program of Chinese Academy of Sciences and the National Natural Science Foundation of China (Project number 11873094). G.X.L. is supported by Yunnan University Grant No. C176220100028. This research has made use of NASA's Astrophysics Data System. This work also made use of Python libraries including Astropy\footnote{\url{https://www.astropy.org/}} \citep{2013A&A...558A..33A}, NumPy\footnote{\url{https://www.numpy.org/}} \citep{5725236}, SciPy\footnote{\url{https://www.scipy.org/}} \citep{jones2001scipy}, Matplotlib\footnote{\url{https://matplotlib.org/}} \citep{Hunter:2007}, APLpy \citep{2012ascl.soft08017R}, seaborn\footnote{\url{https://seaborn.pydata.org/}}, Glue \citep{2015ASPC..495..101B,2017zndo...1237692R}, emcee \citep{2013PASP..125..306F}, and corner.py \citep{corner}. The Overleaf online editor\footnote{\url{www.overleaf.com}} was used to prepare this manuscript. A 3D view of the C$^{18}$O (2--1) position-position-velocity cube can be found in https://gongyan2444.github.io/gallery.html. We thank the anonymous referee for the insightful comments that improve this manuscript.


\begin{appendix}
\section{Zoom-in self-absorption spectra}\label{sec.zoom}
\begin{figure}[!htbp]
\centering
\includegraphics[width = 0.45 \textwidth]{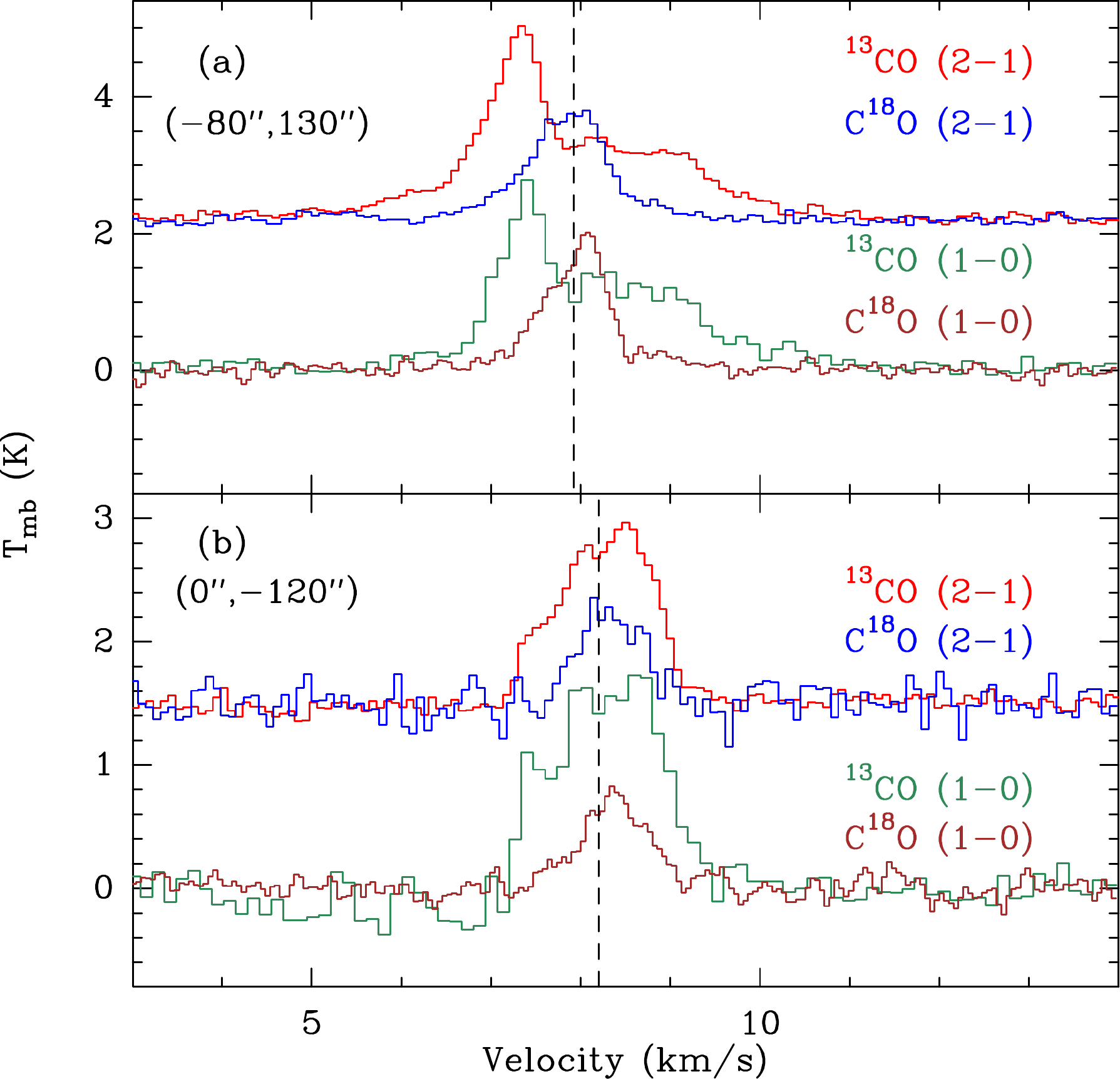}
\caption{{Detailed view on the spectra toward two selected positions, the offsets of which are indicated in the upper left corner. For the origin of the coordinate system, see the caption to Fig.~\ref{Fig:m0}. In each panel, the dashed line marks the velocity of the absorption dip seen in the $^{13}$CO spectra, highlighting that the velocity of the dip varies with position.}\label{Fig:zoom_abs}}
\end{figure}

Here, we present a closer view onto the $^{13}$CO and C$^{18}$O spectra toward two positions in Fig.~\ref{Fig:zoom_abs}. Absorption dips are visible in the two lowest-$J$ $^{13}$CO transitions. The dips appear at 7.92~\kms\,at the offset ($-$80\arcsec, 130\arcsec) and 8.20~\kms\,at the offset (0\arcsec, $-$120\arcsec). The velocity difference corresponds to 8 channels in the $^{13}$CO (1--0) spectra and 4 channels in the $^{13}$CO (2--1) spectra. This demonstrates that the absorption features are not due to contamination in the off position because such a contamination would  cause the dip to occur at the same LSR velocity toward all observed positions.

\section{Additional material for the identified molecular outflow}\label{append.outflow}
\begin{figure*}[!htbp]
\centering
\includegraphics[width = 0.95 \textwidth]{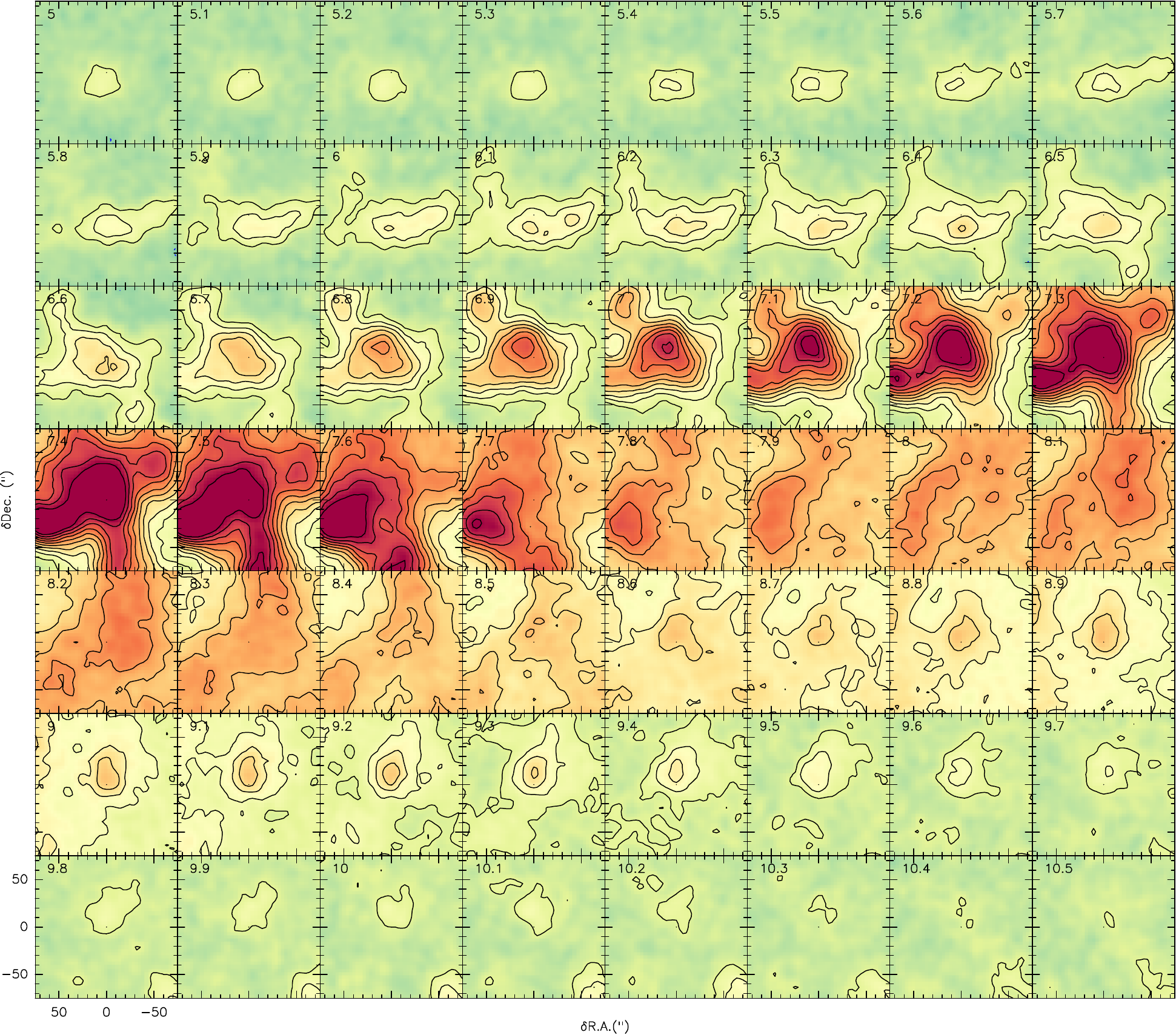}
\caption{{$^{13}$CO (2--1) channel map of the molecular outflow driven by Ser-emb 28. The LSR velocity is indicated in the top left corner of each panel. The contours start at 0.2 K ($3\sigma$), and increase by factors of 2. The offset (0,0) corresponds to the position of Ser-emb 28. The displayed area is the same as Fig.~\ref{Fig:outflow}b.}\label{Fig:cmap}}
\end{figure*}

\begin{figure}[!htbp]
\centering
\includegraphics[width = 0.45 \textwidth]{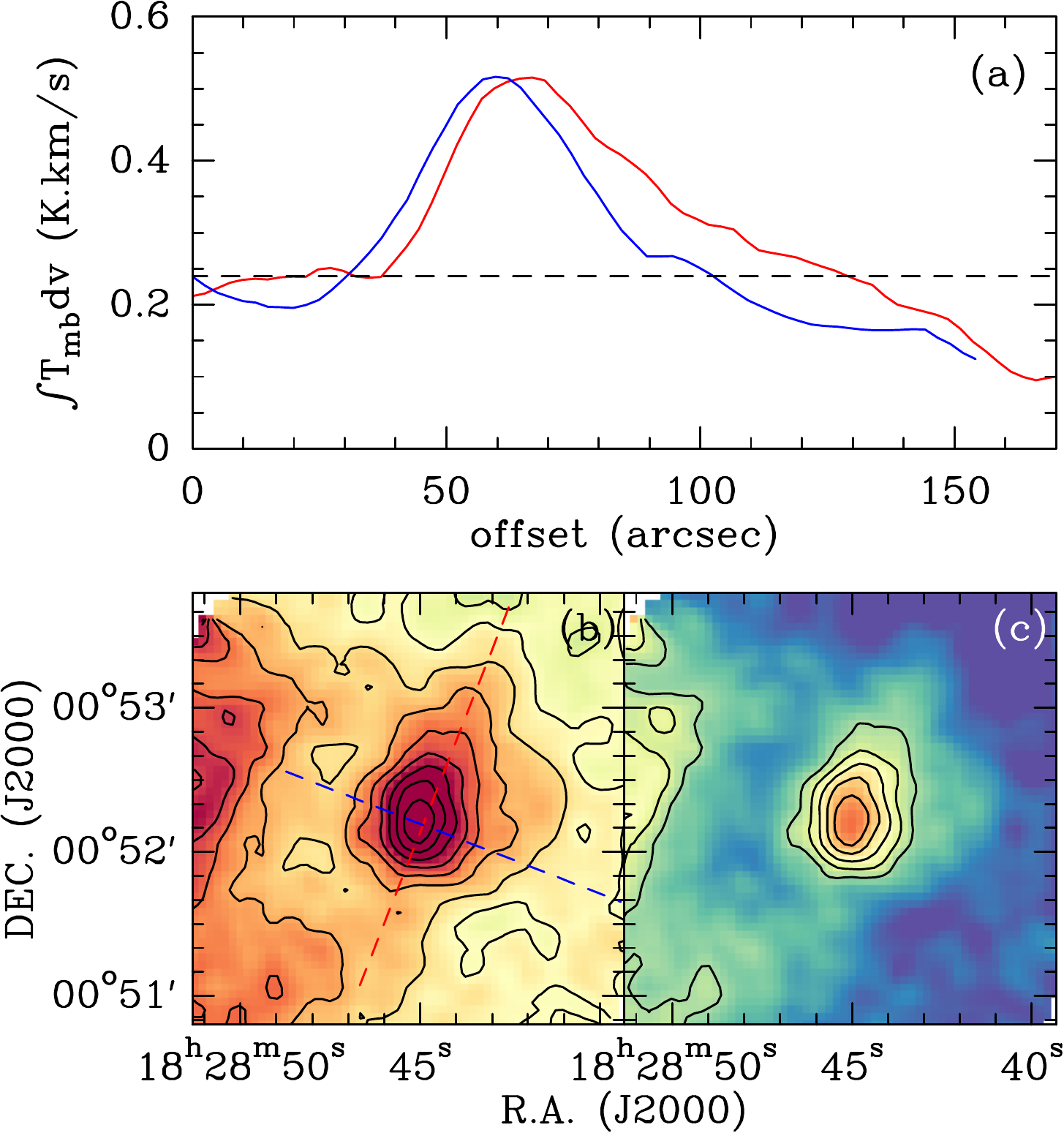}
\caption{{(a) Integrated intensity profiles across the two cuts indicated by the two dashed lines in Fig.~\ref{Fig:out2}b. The offsets increase from south to north for the red line, and from east to west for the blue line. (b) Intensity map integrated from 8.8~\kms\,to 9.4~\kms. The contours start at 0.05~K~\kms (3$\sigma$), and increase by 0.05~K~\kms. (c) Ambient-cloud-emission-subtracted integrated intensity map.}\label{Fig:out2}}
\end{figure}

Here, we present the channel map of the molecular outflow driven by Ser-emb 28 which is shown in Fig.~\ref{Fig:cmap}. Based on this map, we are able to distinguish the outflow and ambient cloud emission at velocities of $\varv_{\rm lsr}\leq$5.5~\kms and $\varv_{\rm lsr}\geq$9.4~\kms. We notice that the compact emission in the range of 8.8--9.3~\kms\,can also arise from the molecular outflow. We investigated the integrated intensity profile (see Fig.~\ref{Fig:out2}a) in the integrated intensity map (see Fig.~\ref{Fig:out2}b). The intensity profiles suggest the presence of a compact component superimposed over extended ambient emission. We found that the ambient emission is characterized by a nearly constant integrated intensity of 0.24~K~\kms\,for each pixel. Subtracting the constant value of 0.24~K~\kms\,from Fig.~\ref{Fig:out2}b, we obtained the residual map shown in Fig.~\ref{Fig:out2}c which was then used to estimate the outflow contribution in the velocity range of 8.8--9.3~\kms. The results are shown as the redshifted-2 component in Table~\ref{Tab:outflow}. We found that these values are not negligible in terms of mass, momentum, and kinetic energy, so its contribution is also taken into account when calculating the overall outflow properties.

\section{nonthermal velocity dispersions with opacity correction}\label{append.a}
\begin{figure}[!htbp]
\centering
\includegraphics[width = 0.45 \textwidth]{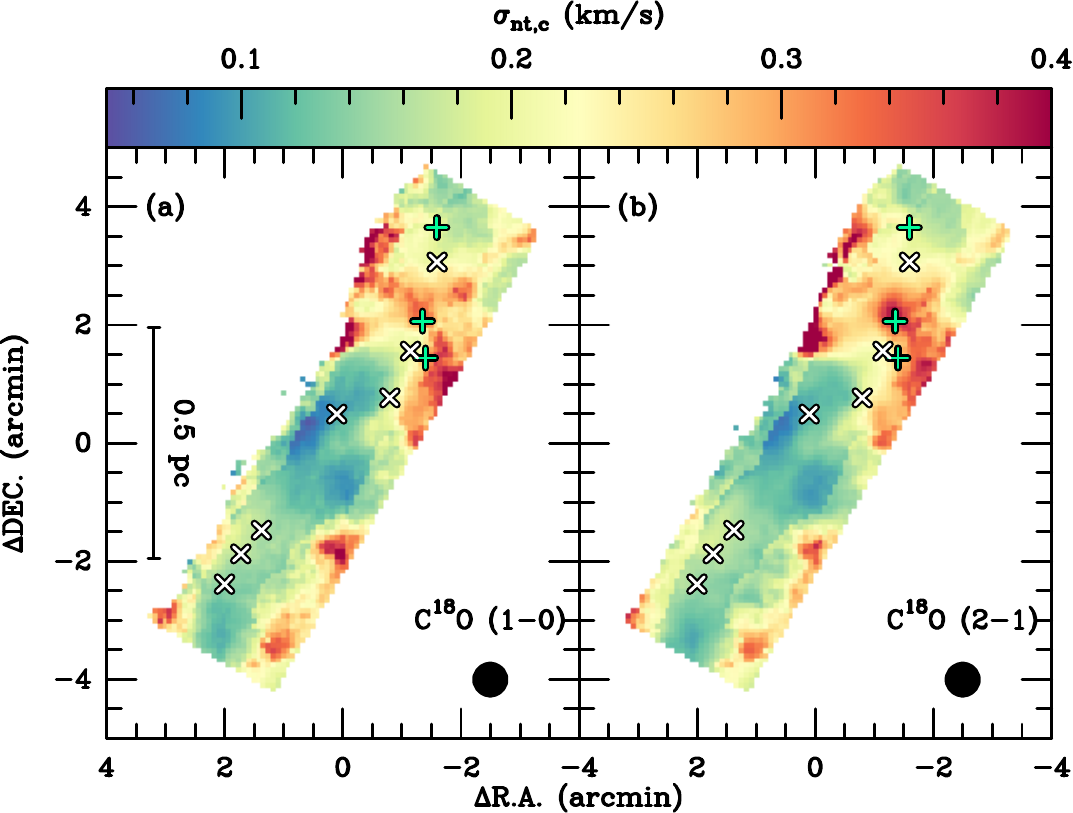}
\caption{{Opacity-corrected nonthermal velocity dispersion maps derived from C$^{18}$O (1--0) and C$^{18}$O (2--1) data. The beam size is shown in the lower right corner of each panel. In both panels, the (0, 0) offset corresponds to $\alpha_{\rm J2000}$=18$^{\rm h}$28$^{\rm m}$50$\rlap{.}^{\rm s}$4, $\delta_{\rm J2000}$=00$^{\circ}$49$^{\prime}$58$\rlap{.}^{\prime \prime}$72, the three green pluses show the positions of the three embedded YSOs, emb10, emb16, and Ser-emb 28 \citep{2009ApJ...692..973E}, and the white crosses mark the positions of the seven 1.1 mm dust cores \citep{2007ApJ...666..982E}. }\label{Fig:nt}}
\end{figure}

As mentioned in Sect.~\ref{sec.kine}, the observed C$^{18}$O line widths are affected by opacity broadening, while the signal-to-noise ratios for C$^{17}$O lines are relatively low. In order to derive the intrinsic velocity dispersion, we need to correct the C$^{18}$O line width for the opacity broadening effects. Following previous studies \citep[e.g.,][]{1979ApJ...231..720P,2016A&A...591A.104H}, the opacity broadening effect can be expressed as:
\begin{equation}\label{f.tauc}
 \frac{\Delta \varv}{\Delta \varv_{\rm int}} = \frac{1}{\sqrt{\rm ln 2}}[{\rm ln}(\frac{\tau_{0}}{{\rm ln}(\frac{2}{{\rm exp}(-\tau_{0})+1})})]^{1/2}\;,
\end{equation}
where $\Delta \varv$ and $\Delta \varv_{\rm int}$ are the observed and intrinsic line widths, respectively. Based on our results shown in Figs.~\ref{Fig:kine} and \ref{Fig:tau}, we find that the ratio $\frac{\Delta \varv}{\Delta \varv_{\rm int}}$ can become as high as 1.4. Subtracting the thermal contributions from the intrinsic line widths, we can derive the opacity-corrected nonthermal velocity dispersion with the relationship:

\begin{equation}\label{f.nt}
\sigma_{\rm nt,c} = \sqrt{\frac{\Delta \varv_{\rm int}^{2}}{8{\rm ln}\,2} - \frac{{\rm k}T_{\rm kin}}{m_{\rm i}}}\;,
\end{equation}
where $\sqrt{\frac{{\rm k}T_{\rm kin}}{m_{\rm i}}}$ is the thermal velocity dispersion, $\sigma_{\rm th}$; $m_{\rm i}$ is the C$^{18}$O mass weight; k is the Boltzmann constant; and $T_{\rm kin}$ is the kinetic temperature. We assumed the kinetic temperature to be 7~K (see Sect.~\ref{sec.lon}), corresponding to an H$_{2}$ sound speed of $c_{\rm s}$= 0.17~\kms~and a C$^{18}$O thermal dispersion of $\sigma_{\rm th}$=0.04~\kms. Applying Eqs.~(\ref{f.tauc})--(\ref{f.nt}) to our observed line widths (see Fig.~\ref{Fig:kine}), we derived the opacity-corrected nonthermal velocity dispersion map which is shown in Fig.~\ref{Fig:nt}. The velocity dispersions outside the 2$\sigma$ boundary in Fig.~\ref{Fig:tau} are also calculated by neglecting the opacity broadening effects. In the inner region, the opacity-corrected nonthermal velocity dispersions range from 0.07 to 0.40~\kms, corresponding to Mach numbers, $M=\sigma_{\rm nt,c}/c_{\rm s}$, of 0.4--2.4. This suggests that the Serpens filament is more quiescent than previously thought \citep{2018A&A...620A..62G}. On the other hand, the region around emb28 becomes supersonic (i.e., $M>$2) likely due to the outflow feedback (see Fig.~\ref{Fig:outflow}). 

The turbulent energy, $E_{\rm t}$, can be estimated with $E_{\rm t} = \Sigma_{i} \frac{1}{2} m_{i}\sigma_{i,{\rm 3d}}^{2}$, where $m_{i}$ and $\sigma_{i,{\rm 3d}}=\sqrt{3}\sigma_{\rm nt,c}$ are the molecular mass and three-dimenional velocity dispersion of the corresponding pixel, $i$. Based on Fig.~\ref{Fig:nt}, we obtain a total turbulent energy of 1.15$\times 10^{44}$ erg, where the turbulent energies of NW and SE are 7.3$\times 10^{43}$ erg and 4.2$\times 10^{43}$ erg, respectively. The difference in the turbulent energies of NW and SE is 3.1$\times 10^{43}$ erg, which is about six times the outflow kinetic energy from Ser-emb 28 (see Table~\ref{Tab:outflow}). Therefore,  the different levels of turbulence in NW and SE cannot be explained only by the outflow feedback. 

The turbulent dissipation rate, $\dot E_{\rm t}$, is defined as,
\begin{equation}\label{f.dissipation}
{\dot E_{\rm t}} =  E_{\rm t}/t_{\rm diss} \, 
\end{equation}
where $t_{\rm diss}$ is the turbulent dissipation time. Based on \citet{2007ARA&A..45..565M}, the turbulent dissipation time is approximately given by
\begin{equation}\label{f.distime}
t_{\rm diss} \approx 0.5 \frac{d}{\sigma_{\rm los}}\,
\end{equation}
where $d$ is the cloud diameter and $\sigma_{\rm los}$ is the turbulent velocity dispersion along the line of sight. We adopt the filament's FWHM length as the cloud diameter, that is, $d$=0.17 pc, and use the mean $\sigma_{\rm nt,c}$ of 0.2~\kms\,as $\sigma_{\rm los}$. We thus obtain the dissipation time of 0.42~Myr. Based on Eq.~\ref{f.dissipation}, we find that $\dot E_{\rm t}$ is about 8.8$\times 10^{30}$ erg~s$^{-1}$ which is comparable to the outflow mechanical luminosity of 5.6$\times 10^{30}$~erg~s$^{-1}$. This suggests that the outflow feedback could sustain the observed turbulent motions.  

\section{Overlapping hyperfine structure lines of C$^{17}$O}\label{append.b}
C$^{17}$O transitions show hyperfine structure (HFS) splitting because of the nonzero nuclear spin in $^{17}$O. The broadening introduced by velocity dispersion and eventually also by opacity can cause the blending of their HFS components. This will introduce additional uncertainties when we estimate the peak optical depth with Eq.~\ref{f.rad} by assuming that a single component or a combined component make the dominant contributions to the peak.

In order to study this impact, we compute the synthetic C$^{17}$O (1--0) and C$^{17}$O (2--1) spectra with Eq.~\ref{f.rad} (see Sect.~\ref{sec.abund}) to compare the intrinsic peak intensity of a single component or a combined component with the peak intensity of the blended spectra. Assuming a Maxwellian velocity distribution, the opacity can be expressed by a Gaussian distribution:
\begin{equation}\label{f.taug}
\tau(\varv) = \tau_{0}{\rm e}^{-(\varv-\varv_{0})^{2}/2\sigma_{\varv}^{2}}\;,  
\end{equation}
where $\tau_{0}$ and $\varv_{0}$ are the opacity and LSR velocity at the line center, and $\sigma_{\varv}$ is the intrinsic velocity dispersion. For the different HFS lines, their opacities are proportional to their relative line strengths assuming LTE \citep{2005JMoSt.742..215M}. Based on the rest frequency of HFS components \citep{2005JMoSt.742..215M}, we are able to obtain their displaced LSR velocities. $\sigma_{\varv}$ should be the same for different components. Taking C$^{17}$O (1--0) for example, the $F=$3/2--5/2, $F=$7/2--5/2, and $F=$5/2--5/2 lines have opacities of 2/9, 4/9, and 1/3 of the total opacity, while the $F=$3/2--5/2 and $F=$5/2--5/2 lines are displaced by 0.5469741~\kms\,and $-$2.7348705~\kms\,with respect to $F=$7/2--5/2. The observed line widths are determined to be 0.25--0.89~\kms\,with a median of 0.52~\kms\,(see Figs.~\ref{Fig:kine}), suggesting that $\sigma_{\varv}$ should be lower than 0.38~\kms\,in the Serpens filament. We therefore use a $\sigma_{\varv}$ range of 0.10--0.38~\kms\,for our calculations. Since both C$^{17}$O lines are likely to have low opacities, we use a peak total opacity of $\tau_{0}\leq$1 for our test. We are thus able to construct the synthetic C$^{17}$O (1--0) and C$^{17}$O (2--1) spectra with Eq.~\ref{f.rad} where $\varv_{0}$ is assumed to be 0 and $T_{\rm ex}$ is set to be 7~K as the fiducial case. 

\begin{figure*}[!htbp]
\centering
\includegraphics[width = 0.95 \textwidth]{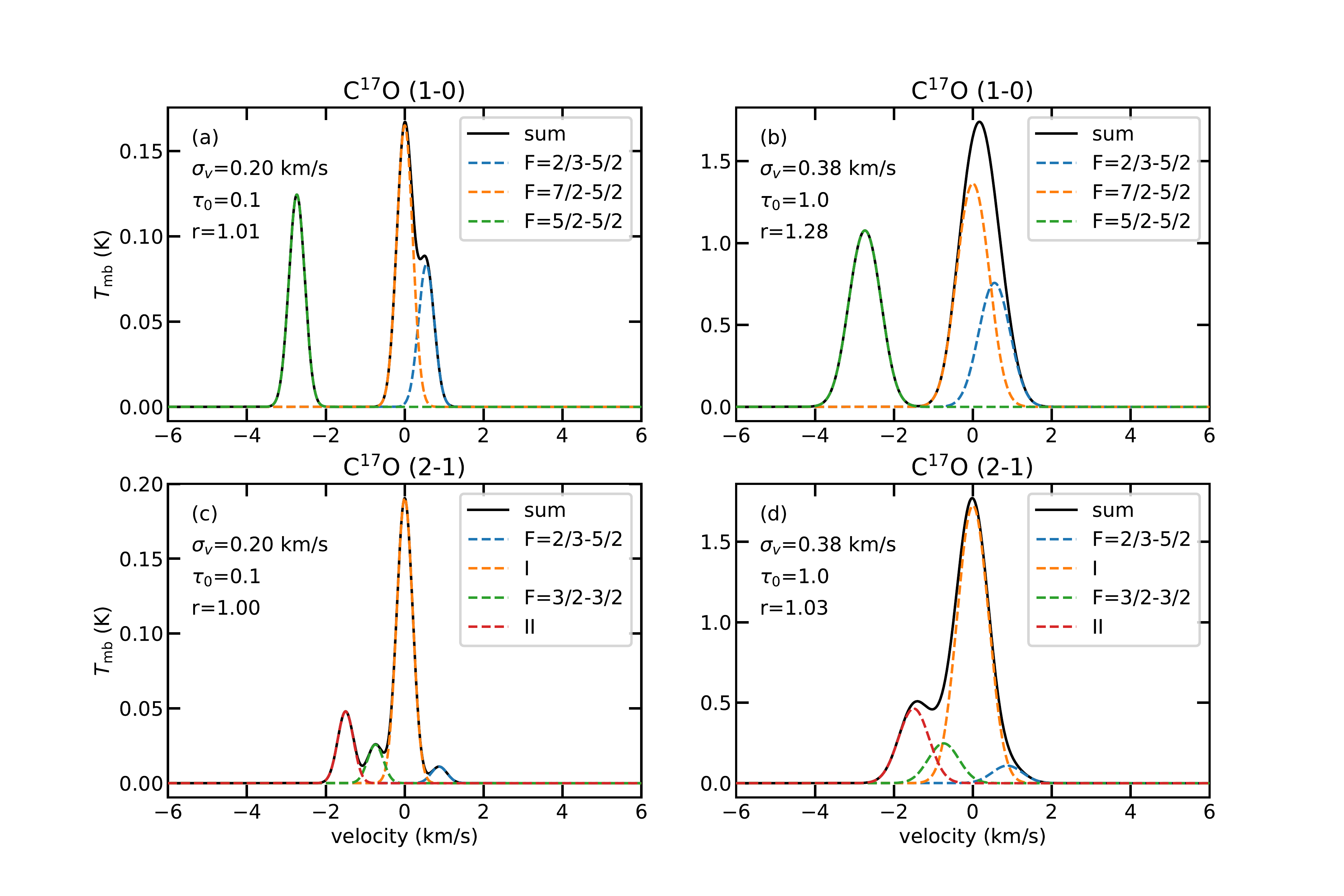}
\caption{{Synthetic C$^{17}$O (1--0) and C$^{17}$O (2--1) spectra with the modeled $\tau_{0}$ and $\sigma_{\varv}$ shown in the upper left corners. For C$^{17}$O (2--1), Group I contains four HFS lines, $F$=5/2--5/2, $F$=7/2--5/2, $F$=9/2--7/2, and $F$=1/2--3/2, while Group II contains three HFS lines, $F$=5/2--7/7, $F$=7/2--7/2, and $F$=5/2--3/2.}\label{Fig:hfs}}
\end{figure*}

The synthetic C$^{17}$O (1--0) and C$^{17}$O (2--1) spectra are shown in Fig.~\ref{Fig:hfs}. Because C$^{17}$O (1--0) and C$^{17}$O (2--1) are dominated by the $F=$7/2--5/2 feature (see Figs.~\ref{Fig:hfs}a--\ref{Fig:hfs}b) and the Group I line (see the Group I line in Figs.~\ref{Fig:hfs}c--\ref{Fig:hfs}d), we use the ratio of the peak intensities of these lines and their corresponding blended lines to assess the difference due to the overlapping HFS lines. We find the ratio increases with increasing $\tau_{0}$ and $\sigma_{\varv}$ for C$^{17}$O (1--0). The highest ratio becomes 0.78 when we use the highest values ($\tau_{0}$=1, $\sigma_{\varv}$=0.38~\kms). This means that the peak intensity can be overestimated by 28\% in this extreme case. For C$^{17}$O (2--1), the ratio does not change much, and the peak intensity is only overestimated by 3\%\,in the extreme case. We also note that the ratio is independent of $T_{\rm ex}$, suggesting that the ratios are robust within our modeling ranges. Therefore, we introduce additional 28\% and 3\% uncertainties for the peak intensities of C$^{17}$O (1--0) and C$^{17}$O (2--1) in studying molecular excitation in Sect.~\ref{sec.abund}.

\end{appendix}

\bibliographystyle{aa}
\bibliography{references}

\end{document}

%% file: obs.tex
\begin{table*}[!hbt]
\caption{Observational parameters related to the molecular lines presented in this work.}\label{Tab:lin}
\normalsize
\centering
\begin{tabular}{cccccccc}
\hline \hline
line             & Frequency         & $E_{\rm u}/k$      & $\theta_{\rm beam}$    & $\delta \varv$     & $\sigma$ & mapping area &  telescope  \\ 
                 & (GHz)             & (K)              &  (\arcsec)           & (\kms)             &   (mK)   & (\arcmin$\times$\arcmin) &             \\ 
(1)              & (2)               & (3)              & (4)                  & (5)                &  (6)     & (7) & (8)         \\
\hline
C$^{18}$O $J=1-0$ & 109.7821734(63)         & 5.3   & 25   & 0.05 &  92           & 2.5$\times$9 & IRAM-30 m        \\
$^{13}$CO $J=1-0$ & 110.2013543(1)         & 5.3   & 25   & 0.14 &  92           & 2.5$\times$9 & IRAM-30 m        \\   
C$^{17}$O $J=1-0$ & 112.358982(15)         & 5.4   & 25   & 0.14 & 100           & 2.5$\times$9 & IRAM-30 m        \\
C$^{18}$O $J=2-1$ & 219.5603541(15)         & 15.8  & 32   & 0.08 &  65           & 3$\times$9   & APEX-12 m        \\
$^{13}$CO $J=2-1$ & 220.3986842(1)         & 15.9  & 32   & 0.08 &  59           & 3$\times$9   & APEX-12 m        \\
C$^{17}$O $J=2-1$ & 224.714187(80)         & 16.2  & 32   & 0.08 &  65           & 3$\times$9   & APEX-12 m        \\
\hline
\end{tabular}
\tablefoot{(1) Observed transition. (2) Rest frequency taken from the Cologne Database for Molecular Spectroscopy \citep[CDMS\footnote{https://zeus.ph1.uni-koeln.de/cdms},][]{2005JMoSt.742..215M}. Uncertainties in the last digits are given in parentheses. For the C$^{17}$O lines, only the frequency of the strongest HFS component is listed. (3) Upper energy level of the observed transition. (4) HPBW. (5) Channel widths in units of velocity. (6) RMS noise level. (7) Size of the mapped area. (8) Telescope.}
\normalsize
\end{table*}

%% file: outflow.tex
\begin{table*}[!hbt]
\caption{Physical parameters of the outflow associated with Ser-emb28.}\label{Tab:outflow}
\small
\centering
\begin{tabular}{cccccccccc}
\hline \hline
lobe        & [$\varv_{\rm min}$, $\varv_{\rm max}$] &  size & $M$         & $P$               & $E$ &  $t$  & $\dot M_{\rm out}$ & $F_{\rm m}$  &  $L_{\rm m}$ \\
            &  (\kms)              &  (pc)   & ($M_{\odot}$) & ($M_{\odot}$~\kms)& (erg) & (yr) & ($M_{\odot}$yr$^{-1}$)  & ($M_{\odot}$~\kms~yr$^{-1}$)  & ($L_{\odot}$)            \\ 
(1)              & (2)               &  (3)                  & (4)                &  (5)     & (6) & (7)  & (8) & (9)   & (10)    \\
\hline
blueshifted & [2, 5.5]    & 0.11   & 0.011 &  0.05  & 2.3$\times 10^{42}$ & 3.4$\times 10^{4}$ & 3.3$\times 10^{-7}$ & 1.5$\times 10^{-6}$ & 5.4$\times 10^{-4}$     \\
redshifted-1  & [9.4, 12]   &  0.09   & 0.028 &  0.09  & 2.9$\times 10^{42}$ & 2.3$\times 10^{4}$ & 1.3$\times 10^{-6}$ & 4.0$\times 10^{-6}$ & 1.0$\times 10^{-3}$     \\
redshifted-2$^{(*)}$  & [8.8, 9.4]  &  0.09   & 0.009 &  0.01  & 2.5$\times 10^{41}$ &   & 3.8$\times 10^{-7}$ & 6.6$\times 10^{-7}$ & 9.0$\times 10^{-5}$     \\
total       &           &       & 0.049  &  0.16  & 5.4$\times 10^{42}$ &                      & 1.9$\times 10^{-6}$ & 5.5$\times 10^{-6}$ &  1.7$\times 10^{-3}$     \\
\hline
\end{tabular}
\tablefoot{(1) Corresponding lobe. (2) LSR velocity range of the wing emission. (3) Length of the lobe in the plane of the sky. (4) Mass. (5) Momentum. (6) Kinetic energy. (7) Dynamical timescale. (8) Mass entrainment rate. (9) Mechanical force (10) Mechanical luminosity. (*) This redshifted component is derived from the ambient-cloud-emission-subtracted integrated intensity
map (details are presented in Appendix~\ref{append.outflow}).}
\normalsize
\end{table*}